\documentclass[12pt, oneside]{article}   
\pdfoutput=1
\usepackage[margin = 1in]{geometry}              
\usepackage[authoryear]{natbib}

\usepackage[parfill]{parskip}    		
\usepackage{graphicx}
\graphicspath{{./}}
\usepackage{multicol}

\usepackage{xcolor}
\usepackage{pagecolor}
\pagecolor{white}

\usepackage{amssymb}
\usepackage{amsmath}
\usepackage{amsthm}

\newtheorem{remark}{Remark}
\newtheorem{lemma}{Lemma}

\usepackage{mathtools}
\usepackage{enumitem}
\usepackage{xfrac}
\usepackage{bm}
\usepackage{titling}
\usepackage{hanging}
\usepackage{setspace}
\usepackage[bottom]{footmisc}
\usepackage{afterpage}
\usepackage{bbm}
\usepackage{enumitem}
\usepackage[hidelinks]{hyperref}

\usepackage{colortbl}
\usepackage{rotating}
\usepackage{booktabs}
\usepackage{floatrow}
\usepackage{bigstrut}
\usepackage{footmisc}
\renewcommand{\thefootnote}{\fnsymbol{footnote}}

\newcounter{daggerfootnote}

\usepackage{etoolbox}
\AtBeginEnvironment{quote}{\par\singlespacing\small}

\usepackage[labelfont=bf, textfont=bf]{caption}

\usepackage{nicematrix}

\usepackage[utf8]{inputenc}
\usepackage[english]{babel}
\newcommand{\LP}{\mathbb{L}}

\newcommand{\indep}{\perp \!\!\! \perp}

\newtheoremstyle{main}%
{3pt}%
{3pt}%
{\itshape}
{}
{\bfseries}
{}
{.5em}%
{}%

\theoremstyle{main}
\newtheorem{assumption}{Assumption}
\newtheorem{proposition}{Proposition}

\DeclareMathOperator*{\argmin}{argmin}

\allowdisplaybreaks
\usepackage{siunitx}

\usepackage{pdflscape}
\usepackage{csquotes}

\begin{document}
\pagestyle{plain}

\thispagestyle{empty}

\setcounter{footnote}{0}

\begin{titlepage}

\noindent \begin{center}
~\\

\Huge{}{Supercompliers\footnote[1]{We thank Josh Angrist, Eric Auerbach, David Card, Amy Finkelstein, Toru Kitagawa, David S. Lee, Doug Miller, Jon Roth, Steve Pischke, and participants in the Princeton University Industrial Relations Section Centennial Symposium and the Stata Virtual Symposium for helpful discussions. Fiona Qiu provided excellent research assistance. Disclaimers: Any opinions and conclusions expressed herein are those of the authors and do not reflect the views of the Joint Committee on Taxation or any Member of Congress. Eng performed this work prior to joining the Internal Revenue Service. All views and opinions expressed herein do not represent the Internal Revenue Service.}}
\par\end{center}

\begin{singlespace}
\begin{center}
{\large{}Matthew L. Comey, Joint Committee on Taxation}\\
{\large{}Amanda R. Eng, Internal Revenue Service}\\
{\large{}Pauline Leung, Cornell University}\\
{\large{}Zhuan Pei, Cornell University and IZA}\footnote[2]{Comey: \href{matthew.comey@jct.gov}{matthew.comey@jct.gov}; Eng: \href{amanda.r.eng@irs.gov}{amanda.r.eng@irs.gov}; Leung: \href{pleung@cornell.edu}{pleung@cornell.edu}; Pei: \href{zhuan.pei@cornell.edu}{zhuan.pei@cornell.edu}}\\
\end{center}
\end{singlespace}

\begin{center}
{\large{}December 2024}{\large\par}
\par\end{center}



\begin{abstract}
\begin{singlespace}

In a binary-treatment instrumental variable framework, we define supercompliers as the subpopulation whose treatment take-up positively responds to eligibility and whose outcome positively responds to take-up. Supercompliers are the only subpopulation to benefit from treatment eligibility and, hence, are important for policy. We provide tools to characterize supercompliers under a set of jointly testable assumptions. Specifically, we require standard assumptions from the local average treatment effect literature plus an outcome monotonicity assumption. Estimation and inference can be conducted with instrumental variable regression. In two job-training experiments, we demonstrate our machinery’s utility, particularly in incorporating social welfare weights into marginal-value-of-public-funds analysis.

\bigskip{}

 \bigskip{}


\end{singlespace}
\end{abstract}

\end{titlepage}

\renewcommand{\thefootnote}{\arabic{footnote}}

\section{Introduction}

Seminal studies by \citet{ImbensandAngrist1994}, \citet{AngristandImbens1995}, and  \citet{Angristetal1996JASA} establish the now well-known local average treatment effect (LATE) interpretation of an estimand from an instrumental variable regression. In the canonical setting with a binary instrument and binary treatment, the LATE is the average treatment effect for individuals who comply with treatment assignment (i.e., compliers). Unsurprisingly, methodological interest in describing compliers followed (\citealt{ImbensandRubin1997}; \citealt{Abadie2003}), and estimation of complier characteristics, popularized by \citet{AngristandPischke2009}, is now widely implemented in empirical studies.\footnote{Recent examples include \citet{Borghansetal2014AEJPo}, \citet{Dahletal2014QJE}, \citet{Dobbieetal2018AER}, \citet{FinkNoto2019}, \citet{OussStevenson2021}, and \citet{AganetalforthcomingQJE}.} Describing the characteristics of compliers provides a direct answer to the question: who are induced to take up treatment when eligible? The exercise is informative of the external validity of the identified treatment effect. As \citet{AngristandPischke2009} note: ``if the compliant subpopulation is similar to other populations of interest, the case for extrapolating estimated causal effects to these other populations is stronger."

In this paper, we extend the complier literature and devise tools to provide a direct answer to a different but related question: who benefit from gaining treatment eligibility? This population consists of those whose treatment take-up responds positively to treatment eligibility \textit{and} whose outcome improves following treatment take-up. In other words, it is the subset of compliers for whom treatment improves the outcome. We term this subpopulation ``supercompliers.''

Supercompliers are a key building block of a LATE. Under a set of testable assumptions and given a binary outcome, we show that the supercomplier population share is the intent-to-treat effect. And since a LATE is the ratio of intent-to-treat and first stage effects, it is equivalent to the supercomplier population share as a fraction of the complier population share. Put simply, the prevalence of supercompliers drives the LATE as we know it. Therefore, learning about supercompliers can be even more informative than learning about the broader complier subpopulation. In the case where supercomplier characteristics differ from complier characteristics, it would be more effective to target treatment eligibility to an external population similar to the supercompliers. 

As with compliers, supercompliers cannot be directly observed. However, we show that their characteristics are identified with expressions analogous to those of complier characteristics. Furthermore, our identification result gives rise to estimators for supercomplier average and distributional characteristics that can be easily implemented with standard instrumental variable regressions. Additionally, in our exploration of various estimators, we show that the plug-in complier characteristics estimators used in the literature can be equivalently implemented using simple instrumental variable regression. Finally, we provide results for the identification and estimation of characteristics for compliers whose outcomes are unaffected by treatment. 

Our key results follow from standard LATE assumptions along with an additional assumption that we call outcome monotonicity. Outcome monotonicity is the outcome counterpart of the treatment monotonicity assumption in the LATE framework. It states that receiving treatment either does not affect the outcome or changes it in a single direction. In other words, treatment does not improve the outcome for some while harming others. 

While outcome monotonicity is not universally tenable, it is plausible in many situations. It is reasonable to assume, for example, that a deworming drug does not lower a child's weight (\citealt{AndrewsandShapiro2021}). More broadly, Institutional Review Board approval for experiments is only granted conditional on demonstrated minimal risks of harm to participants, supporting the assumption of nonnegative treatment effects. Other studies have likewise imposed outcome monotonicity restrictions (e.g., \citealt{Manski1997}, \citealt{ManskiPepper2000}, and \citealt{Lee2009}), highlighting their applicability across a range of settings. 

In other situations, there is ambiguity as to whether outcome monotonicity holds. Hence, it is important that we be able to assess the validity of our identifying assumptions and understand the consequence of violation. To that end, we first extend the seminal work by \citet{Kitagawa2015} to characterize sharp testable implications under the standard LATE assumptions and outcome monotonicity, and we develop an easy-to-implement joint test. Then, to understand the consequence of outcome monotonicity violations, we show that the bias in our estimand is linear in the share of the complier subpopulation harmed by treatment take-up. This result implies that the bias is likely small when few individuals violate the outcome monotonicity assumption. 

Much of our econometric discussion centers around a binary outcome. However, we do not view this focus as restrictive. First, binary outcomes such as employment, welfare receipt, and education credential attainment are quite common in economic research. Second, as argued by \citet{BalkeandPearl1997JASA} who also focus on binary outcomes, we can easily transform a multi-valued or continuous outcome into a binary variable by categorizing its value as ``high'' or ``low.'' Finally, our framework is still interpretable even when the outcome is not binary. Regardless of the distribution of the outcome variable, our estimand identifies a weighted average of supercomplier characteristics, where the weight on each individual is proportional to her treatment effect.

We view analysis of supercomplier characteristics as complementary to subsample heterogeneity analysis of a treatment effect. While conventional heterogeneity analysis asks which subgroups benefit more, our supercomplier analysis collects all beneficiaries and profiles them. Characterizing supercompliers also mirrors the increasingly common practice of describing compliers, which researchers have accepted as a natural complement to subsample heterogeneity analysis of treatment take-up. 

In addition to describing beneficiaries, the supercomplier framework can also be used to enrich marginal value of public funds (MVPF; \citealt{HendrenSprungKeyser2020}) analyses, which have recently become commonplace in empirical economic and policy studies. An input into the MVPF is the social willingness to pay (WTP) for a policy, which is often quantified by the LATE of the policy on some outcome of interest (e.g., earnings). Although social WTP (and therefore MVPF) should be calculated by taking a weighted sum of treatment effects, with weights reflecting the planner's redistributive preferences, it is not feasible in practice when only one LATE is reported. We show, however, that when social welfare weights depend linearly on observed characteristics, a weighted MVPF can be computed using only information on supercomplier characteristics and the reported LATE. A key advantage of this approach is that access to the microdata may not be necessary for recomputing an MVPF with different social welfare functions.

We illustrate our proposed method using data from two well-known randomized experiments on job training. The first is the National Job Corps Study, where eligible youths were randomly assigned access to an intensive residential education and job training program (\citealt{Schochetetal2008}). The second is the National Job Training Partnership Act (JTPA) Study, where economically disadvantaged adults were randomly assigned access to Title II-A programs under the JTPA (\citealt{Bloom_etal1997}). We find that, for labor market outcomes, the Job Corps supercompliers are more advantaged relative to the study population. In contrast, among the group that experienced the largest impact from JTPA---adult women---supercompliers are less relatively advantaged. For a social planner who favors redistribution and places greater welfare weight on the more disadvantaged, our results suggest that the MVPF is lower  for Job Corps and higher for JTPA. 

Related theoretical results were independently developed in exemplary research by \citet{Yu2024}.\footnote{We became aware of this paper in November 2024. The first version of \citet{Yu2024} was circulated in November 2022 as a job market paper. We first posted ours on arXiv in December 2022.} \citet{Yu2024} focuses on the political persuasion literature and interprets outcome-response type profiling in that context. We discuss a range of statistical issues not considered in \citet{Yu2024}, such as interpretation of estimands under stratified randomization, connections between various complier estimators in the literature, identification in the presence of a mediator, and estimation of supercomplier characteristics quantiles. The most important and substantive difference is our emphasis on the interpretability of supercomplier characteristics with a non-binary outcome and the nexus to MVPF analysis. This emphasis significantly broadens the applicability of our framework.

\section{Identification and Estimation\label{sec:theory}}

\subsection{Theoretical Framework and Identification\label{subsec:identification}}

We begin with an extension of the \citet{Rubin1974} potential outcomes framework with a binary instrument $Z \in \{0,1\}$, binary treatment $D \in \{0,1\}$, and binary outcome $Y \in \{0,1\}$. Let $D_z$ represent the potential treatment status for an individual when she is assigned the instrument value $z$. Let $Y_{zd}$ represent a potential outcome for $Z=z$ and $D=d$. With these notations, where each individual's $D$ and $Y$ do not depend on the $Z$ and $D$ of other individuals, we implicitly assume the stable unit treatment value assumption (e.g., \citealt{Cox1958}). 

Throughout this paper, we also maintain the other standard assumptions for the identification of a local average treatment effect (LATE) by \citet{Angristetal1996JASA}. 
\begin{assumption}[IV] \label{A:IV}
\leavevmode
\begin{enumerate}[label={\arabic*.}, align=left]
\item Random Assignment: $(Y_{00}, Y_{01}, Y_{10}, Y_{11}, D_0, D_1) \indep Z$ and $0 < \Pr(Z=1) < 1$. 
\item Exclusion: $\Pr(Y_{1d} = Y_{0d}) = 1 \text{ for } d\in\{0,1\}.$
\item Treatment Monotonicity: $\Pr(D_1 \geq D_0) = 1.$
\item First Stage: $\Pr(D_1 = 1) > \Pr(D_0 = 1).$
\end{enumerate}
\end{assumption}

With Assumptions \ref{A:IV}.1 and \ref{A:IV}.2 along with binary $D$ and $Y$, we can partition the population into 16 unobserved subpopulations or groups based on potential treatment and potential outcome values. These 16 groups also appear in \citet{BalkeandPearl1993} (the working paper version of \citealt{BalkeandPearl1997JASA}) and \citet{ChenFlores2015}, and we refer to this partition as the ``extended principal stratification.'' First, we categorize individuals by how treatment $D$ responds to assignment $Z$. This corresponds to the well-known principal strata (\citealt{FrangakisandRubin2002}) in the LATE context, comprising ``\underline{a}lways takers,'' ``\underline{n}ever takers,'' ``\underline{c}ompliers,'' and ``de\underline{f}iers.'' Second, the exclusion assumption allows us to extend the principal stratification to incorporate how outcome $Y$ responds to treatment $D$ using the same taxonomy. Correspondingly, we define $Y_d \equiv Y_{zd}$ for $z \in \{0,1\}$ to simplify notation. We index each of the 16 groups $G$ by the pair $to$, where $t \in \{a,n,c,f\}$ indexes the four strata based on treatment response and $o \in \{a,n,c,f\}$ indexes the four strata based on outcome response. For each value of $G$, all individuals have the same vector $(D_0,D_1,Y_0,Y_1)$. We fully define the 16 groups in Table \ref{tab:sub}. Supercompliers correspond to the subpopulation $G = cc$, where $D_1 > D_0$ and $Y_1 > Y_0$. 
Assumptions \ref{A:IV}.3 and \ref{A:IV}.4 place restrictions on the shares of these subpopulations. Most importantly, Assumption \ref{A:IV}.3 rules out all treatment defiers, i.e. $G \in \{fa, fn, fc, ff \}$ (for clarity, we use the word ``treatment'' as a qualifier when referring to the conventionally defined always takers, never takers, compliers, and defiers). Together, the two assumptions also require a nonzero share of treatment compliers. Next, we impose an additional assumption that further restricts subpopulation shares from Table \ref{tab:sub}.
\begin{assumption}[Outcome Monotonicity and Reduced Form] \label{A:OM}
\leavevmode
\begin{enumerate}[label={\arabic*.}, align=left]
\item Outcome Monotonicity: $\Pr(Y_1 \geq Y_0 ) = 1.$
\item Reduced Form: $\Pr(G = cc) \equiv \Pr(Y_1 > Y_0, D_1 > D_0)> 0.$
\end{enumerate}
\end{assumption}
Assumption \ref{A:OM} is the outcome analog of Assumptions \ref{A:IV}.3 and \ref{A:IV}.4. Assumption \ref{A:OM}.1 rules out the existence of all outcome defiers (i.e., $G \in \{af, nf, cf, ff\}$). Together, Assumption \ref{A:IV}.3 and Assumption \ref{A:OM}.1 rule out 7 of the 16 groups, and the 9 remaining groups are bolded in Table \ref{tab:sub} (in contrast, \citealt{BalkeandPearl1997JASA} do not rule out groups with monotonicity assumptions but obtain partial identification of the average treatment effect instead). While Assumption \ref{A:IV}.4 implies a nonzero population share of treatment compliers and consequently a nonzero first stage, Assumption \ref{A:OM}.2 requires a nonzero population share of supercompliers, which implies a nonzero intent-to-treat effect (or reduced form) as shown in Lemma \ref{lemma:decomp} below. (Along with Assumptions \ref{A:IV}.1-\ref{A:IV}.3, Assumption \ref{A:OM}.2 implies Assumption \ref{A:IV}.4.)

\begin{lemma}
\label{lemma:decomp}
Under Assumptions \ref{A:IV} and \ref{A:OM}.1, the supercomplier share is identified by:
\begin{equation*}
    \Pr(G=cc)=E[Y|Z=1]-E[Y|Z=0].
\end{equation*}
\end{lemma}

All proofs are in the Appendix.

Lemma \ref{lemma:decomp} states that the share of the supercompliers is identified by the reduced form. This corresponds naturally to the well-known result that the share of treatment compliers is identified by the first stage. Since the reduced form is the product of the local average treatment effect and the first stage, the LATE (given a binary outcome) is simply the share of supercompliers as a fraction of the treatment compliers. This result is intuitive: only the outcome compliers have a nonzero treatment effect (their treatment effect is equal to one), while all other admissible treatment compliers have a zero treatment effect. The more supercompliers there are, the higher the LATE.  

In fact, the supercomplier group is the only subpopulation under Assumptions \ref{A:IV} and \ref{A:OM} whose outcome changes with the assignment $Z$. In other words, they are the only ones who benefit from being assigned to the treatment group. As such, learning about supercompliers should be of great importance to policy makers. 

As with compliers, we cannot directly identify members of the supercomplier subpopulation. However, we can identify the distribution of their characteristics using observed data. Let $X$ be a variable determined prior to random assignment (i.e., prior to the realization of $Z$), so it is reasonable to assume that jointly with $(Y_1,Y_0,D_1,D_0)$, $X$ is independent of $Z$ (hereafter we use $X \indep Z$ as a shorthand for this joint independence). For simplicity, we consider the case of a one-dimensional $X$, but many of our results generalize to a covariate vector of any dimension. Let $h$ be a function such that $E[|h(X)|]<\infty$.
\begin{proposition}
\label{prop:SC_Char}
Under Assumptions \ref{A:IV} and \ref{A:OM} and provided that $X \indep Z$, the supercomplier average of $h(X)$ is identified by:
\begin{equation}
\label{eq:sc_char_id_weight}
E[h(X) | G=cc]  =  \frac{1}{RF} E[\pi h(X)],
\end{equation}
where $RF \equiv E[Y | Z=1] - E[Y|Z=0]$, and $\pi \equiv \kappa - (\kappa_0 Y + \kappa_1(1-Y))$ with
\begin{align*}
\kappa &\equiv 1 - \frac{D(1-Z)}{\Pr(Z=0)} - \frac{(1-D)Z}{\Pr(Z=1)} \\
\kappa_0 &\equiv \frac{(1-D)(1-Z)}{\Pr(Z=0)} - \frac{(1-D)Z}{\Pr(Z=1)} \\
\kappa_1 &\equiv \frac{DZ}{\Pr(Z=1)} - \frac{D(1-Z)}{\Pr(Z=0)}.
\end{align*}
The supercomplier average can also be identified by a Wald-type estimand:
\begin{equation}
E[h(X) | G=cc]  = \frac{E[h(X)Y | Z = 1] - E[h(X)Y | Z=0] }{ E[Y | Z=1] - E[Y|Z=0] }.
\label{eq:sc_char_id_wald}
\end{equation}
\end{proposition}
The weights $\kappa$, $\kappa_0$, and $\kappa_1$ are the unconditional counterpart of those defined by \citet{Abadie2003} in his study of compliers. Lemma \ref{lemma:Abadie} in Appendix \ref{subsec:proofs} provides the analog of Theorem 3.1 from \citet{Abadie2003} under our Assumption \ref{A:IV}. While \citet{Abadie2003} assumes the LATE assumptions hold conditional on $X$, our Lemma \ref{lemma:Abadie} shows that all three weights can still be used to identify the $X$ distribution of the compliers when the assumptions hold unconditionally. Lemma \ref{lemma:Abadie} also shows that $\kappa_0$ ($\kappa_1$) can still be used to identify the $Y_0$ ($Y_1$) distribution of the compliers under our assumptions.

\begin{remark}
\normalfont
({\bf Compliers and Supercompliers}) There is a natural parallel between the identification results for supercomplier characteristics above and those for complier characteristics from the literature. While the $\kappa$ weights from \citet{Abadie2003} ``find compliers'' (\citealt{AngristandPischke2009}), our $\pi$ weights find supercompliers. Heuristically, the $\kappa$ weight starts from the population and subtracts the treatment always-takers and never-takers, while the $\pi$ weight starts from the treatment compliers and subtracts the $G=ca$ and $G=cn$ groups. There is a complier analog to equation (\ref{eq:sc_char_id_wald}) as well: \citet{KlineWalters2016} and \citet{MarbachandHangartner2020} show that the complier characteristics can be identified by replacing $Y$ with $D$ in (\ref{eq:sc_char_id_wald}).  
\end{remark}

\begin{remark}
\normalfont
({\bf Perfect Treatment Compliance}) Our estimands from Lemma \ref{lemma:decomp} and Proposition \ref{prop:SC_Char} are applicable to the case where $D=Z$, i.e., the case of perfect compliance with treatment assignment.\footnote{This is the setting \citet{Kowalski2020} considers, although she adheres to the conventional nomenclature and studies ``defiers'' where ``take-up'' is defined using an outcome variable.}  Here, the supercomplier share from Lemma \ref{lemma:decomp} is equal to the population average treatment effect. And the expressions for supercomplier characteristics identification under perfect treatment compliance are isomorphic (i.e., identical up to variable labels) to those for complier characteristics identification under treatment noncompliance.
\end{remark}

\begin{remark}
\label{remark:distribution}
\normalfont
({\bf Identification of the Distributions of Characteristics}) When $h$ is the identity function, Proposition \ref{prop:SC_Char} says that we can identify the average characteristics of the supercompliers. We can also choose $h=1_{[X \leq x]}$ for all $x \in \mathbb{R}$ and identify the entire c.d.f. of $X$ among the supercompliers. More generally, when $X$ is $k$-dimensional, letting $h=1_{[X \leq x]}$ for $x \in \mathbb{R}^k$ identifies the joint distribution of the random vector $X$ among the supercompliers. 
\end{remark}

\begin{remark}
\normalfont
({\bf Share and Characteristics of Other Groups}) We can also identify the shares and characteristics of the other two groups within the unobserved treatment complier population: $G=ca$ and $G=cn$. See Appendix \ref{subsec_app:othergroups} for details.
\end{remark}

\begin{remark}
\label{remark:mediator}
\normalfont
({\bf Adding a Mediator}) The tools we have developed can also shed light on how a binary mediator, $M$, operates on different populations (the causal chain is $Z \to D \to M \to Y$). If we extend the exclusion restriction and monotonicity assumption to cover $M$ (that is, $Y_{zdm}$=$Y_m$ and $M_1>M_0$), our supercomplier estimand simply identifies the characteristics of those with $D_1>D_0,M_1>M_0,Y_1>Y_0$ (the ``superdupercompliers''). In addition, we can identify the shares and characteristics of those with $D_1>D_0,M_1=M_0=m$ and $D_1>D_0,M_1>M_0,Y_1=Y_0=y$ for $m,y=0,1$. See Appendix \ref{subsec_app:mediator_details} for details.
\end{remark}

We can generalize Lemma \ref{lemma:decomp} and Proposition \ref{prop:SC_Char} to accommodate non-binary $Y$, where supercompliers are still defined as those with $D_1>D_0$ and $Y_1>Y_0$ and referred to with the shorthand $G=cc$. The proposition below shows that regardless of the distribution of $Y$, the reduced-form estimand in Lemma \ref{lemma:decomp} identifies the supercomplier population share scaled by the average treatment effect among the supercompliers. Similarly, the Wald estimand in Proposition \ref{prop:SC_Char} identifies a weighted average of supercomplier characteristics, where the weight of each individual is proportional to her treatment effect.

\begin{proposition}
\label{prop:non_binary_Y}
Under Assumptions \ref{A:IV} and \ref{A:OM}, the reduced form identifies the supercomplier share scaled by the supercomplier average treatment effect:
\begin{equation}
\label{eq:share_non_binary_Y}
    \Pr(G=cc)E[Y_{1}-Y_{0}|G=cc]=E[Y|Z=1]-E[Y|Z=0].
\end{equation}
With the additional assumption that $X \indep Z$, the Wald estimand identifies supercomplier characteristics weighted by treatment effect:
\begin{equation}
\label{eq:char_non_binary_Y}
    \frac{E[h(X)(Y_{1}-Y_{0})|G=cc]}{E[Y_{1}-Y_{0}|G=cc]}=\frac{E[h(X)Y|Z=1]-E[h(X)Y|Z=0]}{E[Y|Z=1]-E[Y|Z=0]}.
\end{equation}
\end{proposition}
As we demonstrate in Sections \ref{sec:MVPF} and \ref{sec:empirical}, Proposition \ref{prop:non_binary_Y} substantially widens the applicability of our machinery to different types of outcomes and is essential for our MVPF analysis. 

More broadly, while our discussion here has focused on randomized experiments, our framework can be applied to quasi-experimental research designs by making appropriate adjustments to the underlying assumptions. In a regression discontinuity design, for instance, we can extend our logic and identify supercompliers at the policy threshold under smoothness conditions. 
We now turn to estimation and inference.

\subsection{Estimation and Inference\label{subsec:estimation}}

The Wald-type estimand in (\ref{eq:sc_char_id_wald}) and (\ref{eq:char_non_binary_Y}) can be implemented via a two-stage least squares (2SLS) regression. For example, when $h(X)$ is the identity function---corresponding to the mean characteristics of supercompliers---we can run this regression in Stata as
\begin{equation}
{\tt ivregress \;\; 2sls \;\; XY \;\; (Y \; = \; Z ) \;\; [, \;\; options]}
\label{eq:stata}
\end{equation}
where {\tt XY} is a variable defined as the product of $X$ and $Y$.

However, equation (\ref{eq:stata}) does not estimate an exact sample analog of the Abadie-style equation (\ref{eq:sc_char_id_weight}) from Proposition \ref{prop:SC_Char}. As we show in Appendix \ref{subsec_app:estimator_details}, the estimand from equation (\ref{eq:sc_char_id_weight}) also has a Wald-type representation similar to (\ref{eq:sc_char_id_wald}): 
\begin{equation}
\frac{E[h(X)\{Y-(1-\tau)\} | Z = 1] - E[h(X)\{Y-(1-\tau)\} | Z=0] }{ E[Y-(1-\tau) | Z=1] - E[Y-(1-\tau)|Z=0] },
\label{eq:sc_char_id_weight_wald}
\end{equation}
where $\tau \equiv \Pr(Z=1)$ denotes the proportion of units assigned to treatment. Therefore, the sample analog of (\ref{eq:sc_char_id_weight}) can also be implemented using 2SLS. But we need to first transform $Y$ by subtracting from it the proportion of observations in the control group, and then use this transformed outcome variable in the 2SLS regression (it turns out that we can ignore the sampling variation in estimating $\tau$ when conducting inference). As discussed in Appendix \ref{subsec_app:estimator_details}, neither of the two estimators based on (\ref{eq:sc_char_id_weight}) and (\ref{eq:sc_char_id_wald}) has an asymptotic variance that dominates the other in all data generating processes (DGPs).

Estimators based on complier analogs of both (\ref{eq:sc_char_id_weight}) and (\ref{eq:sc_char_id_wald}) have been used in existing studies, for which $Y$ is replaced by $D$. \citet{AngristandPischke2009} implement the complier analog of (\ref{eq:sc_char_id_weight}). Other empirical studies referenced in the introduction (e.g., \citealt{FinkNoto2019}) use a different estimator, which is equivalent to the complier analog of (\ref{eq:sc_char_id_wald}) (see Appendix \ref{subsec_app:estimator_details} for details). However, these studies do not use 2SLS regression to estimate complier characteristics. Instead, implementation involves assembling separately estimated quantities. In addition, these studies either do not report standard errors on estimated complier characteristics or report bootstrapped standard errors. Our results here imply that inference results can be easily obtained for both estimators using existing Stata commands, including those that account for a weak instrument.\footnote{\label{fn:weakiv}For estimating complier characteristics, a weak instrument has the standard meaning---$Z$ fails to generate sufficient variation in $D$. For supercompliers, we have a weak instrument problem if $Z$ fails to generate sufficient variation in $Y$.}

Recent work by \citet{AngristHullWalters2023} follows and extends the arguments by \citet{Abadie2002} and uses additional estimators to characterize compliers. In addition to the complier analog of (\ref{eq:sc_char_id_wald}), which amounts to using $\kappa_1$ weights, they also consider weighting covariates by $\kappa_0$, equivalent to replacing $Y$ by $1-D$ in (\ref{eq:sc_char_id_wald}). \citet{AngristHullWalters2023} also propose a ``pooled" estimator, which is the average of the $\kappa_1$ and $\kappa_0$ weighted estimators and can be implemented via a stacked regression. While we can easily use the supercomplier analogs of these additional estimators, replacing $Y$ with $1-Y$ seems unnatural when $Y$ is non-binary. Therefore, we simply use the sample analog of the Wald estimand in (\ref{eq:sc_char_id_wald}) and (\ref{eq:char_non_binary_Y}) to estimate supercomplier characteristics, which is akin to the complier IV estimators used by \citet{Alsanetal2024} when the treatment variable is non-binary.

\begin{remark}
\normalfont
({\bf Characteristics Distribution and Quantile Estimation}) As we discussed in Remark \ref{remark:distribution}, we can identify the entire $X$ distribution among supercompliers. It is straightforward to estimate the c.d.f. of $X$: we can replace the dependent variable in (\ref{eq:stata}) with the product of the indicator function $1_{[X \leq x]}$ and $Y$ and estimate a series of 2SLS regressions by varying $x$. To estimate the supercomplier quantiles of $X$, we can minimize a weighted sum of the check function per \citet{BassettandKoenker1982}. However, the same challenge encountered by \citet{Abadieetal2002} in complier quantile estimation is also present here---since the individual $\pi$ weight may be negative, the sample objective function is usually non-convex and is therefore difficult to minimize. \citet{Abadieetal2002} overcome this challenge by using weights conditional on $(D,Y,X)$. Directly applying this strategy to the supercomplier setting does not lead to nonnegative weights, but applying a modified strategy works, in which we use the $\pi$ weights only conditional on $(Y,X)$ or just $X$. See Appendix \ref{subsec_app:quantiles} for details.
\end{remark}

\begin{remark}
\label{rem:cond_indep}
\normalfont
({\bf Conditional Independence and Stratified Randomization}) Our identification and estimation results can be naturally generalized to accommodate cases where independence (Assumption \ref{A:IV}.1) holds conditionally on covariate set $W$---we just need to add $W$ into the conditioning set in Lemma \ref{lemma:decomp} and Proposition \ref{prop:SC_Char}. A common situation that calls for conditional independence is stratified randomized experiments, in which researchers typically include stratum fixed effects in treatment effect regressions (\citealt{BruhnandMcKenzie2009}). It is then natural to also include the stratum fixed effects in the IV regression estimating supercomplier characteristics. In Appendix \ref{subsec_app:cond_indep}, we follow \citet{Blandholetal2022} to show that the resulting population regression coefficient still identifies a non-negatively weighted average of supercomplier characteristics across strata.
\end{remark}

\subsection{The Outcome Monotonicity Assumption\label{subsec:OM}}

While Assumption \ref{A:IV} is standard in the RCT literature, Assumption \ref{A:OM}.1 (outcome monotonicity) warrants more discussion. First, we point out that previous studies have maintained similar assumptions. For example, in their influential studies of treatment effect partial identification, outcome monotonicity is what \citet{Manski1997} and \citet{ManskiPepper2000} refer to as ``monotone treatment response'' and what \citet{Lee2009} refers to simply as ``monotonicity.''  In their motivating examples, outcome monotonicity is taken to mean that the demand curve is weakly downward sloping (\citealt{Manski1997}), that education does not decrease wages (\citealt{ManskiPepper2000}), or that participating in the Job Corps training program does not lower employment (\citealt{Lee2009}).\footnote{\citet{ChenFlores2015} extend \citet{Lee2009} to bound treatment effects under imperfect compliance. Their monotonicity assumption is akin to Jobs Corps \emph{take-up} not lowering employment for the treatment compliers, while  \citet{Lee2009} assumes \emph{assignment} to Jobs Corps does not lower employment. Under treatment monotonicity, these assumptions are equivalent.} Additionally, outcome monotonicity is implicitly assumed in the classic constant parameter endogenous treatment model of \citet{Heckman1978}. 
As with Assumption \ref{A:IV}, the practical plausibility of Assumption \ref{A:OM} depends on the context. For example, it is quite plausible for the relationship between training program participation and subsequent employment or between health insurance coverage and doctor visits to be weakly positive. However, there is more ambiguity concerning the relationship between, say, health insurance coverage and out-of-pocket medical spending. Health insurance may lead to significant savings during emergency room visits, but it may also incentivize healthcare utilization and lead to higher spending.

Given uncertainty in the plausibility of this and the other identifying assumptions, researchers would be prudent to test them. We state such a test below. 

\subsubsection{Assumption Testing\label{subsubsec:test_id}}

We build on and extend results from \citet{Kitagawa2015} and propose a sharp characterization of Assumptions \ref{A:IV}.1-\ref{A:IV}.3 and \ref{A:OM}.1.\footnote{We exclude Assumption \ref{A:IV}.4 (nonzero first-stage) and Assumption \ref{A:OM}.2 (nonzero reduced-form) from the joint test below for two reasons. First, this exclusion is consistent with \citet{Kitagawa2015} who only tests Assumptions \ref{A:IV}.1-\ref{A:IV}.3 and not Assumption \ref{A:IV}.4. Second, testing for nonzero first-stage and reduced-form is straightforward and a must-do in any empirical study.} The resulting test takes the form of a set of inequalities that must jointly hold.\footnote{Note that testing treatment and outcome monotonicity (Assumptions \ref{A:IV}.3 and \ref{A:OM}.1) amounts to testing for the existence of treatment and outcome defiers, respectively. While we do not consider it here, \citet{Kowalski2020} proposes a finite sample test of the existence of outcome defiers (\citealt{Kowalski2020} simply refers to them as defiers) in a perfect compliance framework. Extending her results to accommodate incomplete take-up is an avenue for future research.} Consistent with \citet{Kitagawa2015}, ``sharp" here means that if the inequalities hold, then we can construct a data generating process which satisfies Assumptions \ref{A:IV} and \ref{A:OM} and rationalizes the observed data. Formally,

\begin{proposition}
\label{prop:sharp}
Given the potential outcomes model described in Section 2.1, (i) under Assumptions \ref{A:IV}.1-\ref{A:IV}.3 and \ref{A:OM}.1, the following inequalities hold
\begin{align}
    \Pr(Y = 0, D=1 | Z = 1)-\Pr(Y = 0, D=1 | Z = 0) &\geq 0 \label{prop_sharp_ineq:1} \\
	\Pr(Y = 1, D=0 | Z = 0)-\Pr(Y = 1, D=0 | Z = 1) &\geq 0 \label{prop_sharp_ineq:2} \\
	\Pr(Y = 1 | Z = 1)-\Pr(Y = 1 | Z = 0) &\geq 0 \label{prop_sharp_ineq:3};
\end{align}
(ii) if inequalities (\ref{prop_sharp_ineq:1})-(\ref{prop_sharp_ineq:3}) hold, there exists a joint distribution of $(Y_{11},Y_{10},Y_{01},Y_{00},D_1,D_0,Z)$ that satisfies Assumptions \ref{A:IV}.1-\ref{A:IV}.3 and \ref{A:OM}.1 and induces the observed distribution of $(Y, D, Z)$.
\end{proposition}

Our identification results for the size of the treatment complier subpopulations (Lemma \ref{lemma:decomp} and Proposition \ref{prop:other_groups} in the Appendix) reveal an intuitive interpretation of inequalities (\ref{prop_sharp_ineq:1})-(\ref{prop_sharp_ineq:3}). The left-hand-sides identify the population shares of the $cn$, $ca$, and $cc$ groups, respectively, under our assumptions. Therefore, the testable implication of the identifying assumptions is simply that these quantities are weakly positive, which may not be the case if, for example, the disallowed subpopulation shares are positive.

Our test is related to a result presented in \citet{Machadoetal2019}, who study the identification of the sign of the average treatment effect in a LATE setting. Their Theorem 3.2(ii) establishes a set of inequalities that hold if and only if the standard LATE assumptions are satisfied and the average treatment effect is nonnegative. In our binary outcome setting, the \citet{Machadoetal2019} inequalities turn out to be  identical to those in Proposition \ref{prop:sharp}. While an assumption of a nonnegative average treatment effect is implied by and therefore weaker than our outcome monotonicity assumption, data cannot tell the two apart. Indeed, the proof of Proposition \ref{prop:sharp} reveals that all the information available for evaluating outcome monotonicity is contained within the reduced form estimate (Inequality \ref{prop_sharp_ineq:3}), which translates to a nonnegative average treatment effect.

Testing the inequalities one-by-one is straightforward: Each requires a one-sided test based on a treatment-control comparison. To test all three inequalities jointly, we propose running a ``stacked'' regression, where each stack corresponds to a single inequality. That is: first, create three copies of the data; second, for each individual $i$, define $Y_{i}^{1} = (1-Y_i)D_i$,  $Y_{i}^{2} = Y_i(1-D_i)$, and $Y_{i}^{3} = Y_i$; and third, estimate the regression
\begin{equation}
\label{eq:joint_test}
	Y_{i}^{s} = \phi^{s} + \theta_1 Z_i 1_{[s=1]} + \theta_2 (1-Z_i) 1_{[s=2]} + \theta_3 Z_i 1_{[s=3]} + \varepsilon_{i}^{s},
\end{equation}
where $s$ indexes each stack, $\phi^{s}$ is the stack specific constant, and standard errors are clustered at the individual level. The joint test amounts to testing whether $\theta \equiv \min(\theta_1,\theta_2,\theta_3)$ is nonnegative. Specifically, because the estimators $(\hat{\theta}_1,\hat{\theta}_2,\hat{\theta}_3)$ are asymptotically multivariate normal with a covariance matrix we can consistently estimate, we can simulate the asymptotic distribution of $\hat{\theta}$ under the null. We reject the null hypothesis and conclude violations of the identifying assumptions when $\hat{\theta}$ is less than the 5th percentile in that distribution.

\begin{remark}
\label{rem:OM_test_only}
\normalfont
Inequality (\ref{prop_sharp_ineq:3}) corresponds to testing outcome monotonicity (Assumption \ref{A:OM}.1). If we wish to test outcome monotonicity under the LATE Assumptions (Assumptions \ref{A:IV}.1-\ref{A:IV}.3)---as opposed to testing all these assumptions jointly---we can simply rely on inequality (\ref{prop_sharp_ineq:3}) alone.
\end{remark}

\begin{remark}
\label{rem:generalized_test}
\normalfont
We can generalize the test in Proposition \ref{prop:SC_Char} by removing the binary restriction on $Y$. In Appendix \ref{subsec:generalized_test}, we state a sharp joint test of our identifying assumptions, which allows $Y$ to have an arbitrary distribution. We implement this generalized test in our empirical applications where $Y$ represents earnings, an important outcome in MVPF analyses.
\end{remark}

\begin{remark}
\label{rem:OM_test_cov}
\normalfont
We can incorporate covariates to increase power in the outcome monotonicity test. Specifically, instead of using inequality (\ref{prop_sharp_ineq:3}) to test Assumption 2.1, we check whether the inequality holds when further conditioning on covariates. A simple way to implement this covariate-augmented test is to first discretize the covariate space (if necessary) and then jointly test the conditional analog of inequality (\ref{prop_sharp_ineq:3}) via a stacked regression, in which each stack corresponds to a value of the covariate vector. In Appendix \ref{sec_app:outcome_monotonicity_CTJF}, we provide a proof of concept by showing that this test indeed rejects outcome monotonicity in a well known example by \citet{Bitleretal2006,Bitleretal2017} where Assumption \ref{A:OM}.1 fails. 
\end{remark}

While the result from the exercise in Appendix \ref{sec_app:outcome_monotonicity_CTJF} is reassuring, we acknowledge that our testing procedures cannot detect all violations of the identifying assumptions. By following the same construction as in the proof of \citet{Kitagawa2015}'s Proposition 1.1(ii), we can show that for any observed $(Y,D,Z)$ that satisfies inequalities (\ref{prop_sharp_ineq:1})-(\ref{prop_sharp_ineq:3}), there exists an underlying joint distribution $(Y_{11},Y_{10},Y_{01},Y_{00},D_1,D_0,Z)$ that violates some of the identifying assumptions. If we use inequality (\ref{prop_sharp_ineq:3}) or its extension conditional on covariates as mentioned in Remark \ref{rem:OM_test_cov} to only test outcome monotonicity, we will fail to detect a violation if for every covariate value, the $G=cf$ group (or the ``compfier" group) share is nonzero but is less than the $G=cc$ (supercomplier) share. Finally, like any other statistical validity test, we may lack the precision to detect a violation in a given sample. For all these reasons, we investigate the consequence of outcome monotonicity violations next. 

\subsubsection{Relaxing Outcome Monotonicity\label{subsubsec:relax}}

Proposition \ref{prop:No_OM} below shows what the supercomplier characteristics estimand identifies when we relax outcome monotonicity (Assumption \ref{A:OM}.1). It is analogous to Proposition 3 in \citet{Angristetal1996JASA} on relaxing treatment monotonicity (our Assumption \ref{A:IV}.3).

\begin{proposition}
\label{prop:No_OM}
Under Assumptions \ref{A:IV} and \ref{A:OM}.2 and provided that $X \indep Z$,
\begin{align*}
\frac{E[h(X)Y|Z=1]-E[h(X)Y|Z=0]}{E[Y|Z=1]-E[Y|Z=0]}=&E[h(X)|G=cc]+\\&\xi\left\{ E[h(X)|G=cc]-E[h(X)|G=cf]\right\} ,    
\end{align*}
where we define
\begin{equation*}
\xi \equiv \frac{\Pr(G=cf) }{E[Y|Z=1]-E[Y|Z=0]}.
\end{equation*}
\end{proposition}

Proposition \ref{prop:No_OM} says that when outcome monotonicity does not hold, the estimand for supercomplier characteristics will be biased unless the supercompliers and the compfiers have the same average characteristics. The bias increases linearly with the compfier share, which can be interpreted as the degree of an outcome monotonicity violation. The bias is small when the degree of outcome monotonicity violation is low.

In spite of our discussion here and in Section \ref{subsec:OM}, we acknowledge that needing outcome monotonicity is a drawback of our supercompliers framework relative to subsample heterogeneity analysis, which increasingly relies on machine learning techniques (e.g, \citealt{AtheyandImbens2016}, \citealt{WagerandAthey2018}; see \citealt{Smith2022} for a recent review). But our approach also has several advantages. First, it allows imperfect compliance with treatment, from which machine learning methods often abstract away. Second, it avoids any approximation of the conditional expectation of potential outcomes or treatment effects with covariates. Consequently, our result does not depend on the choice of a particular machine learning algorithm or tuning parameters therein (see, for example, \citealt{Knausetal2021EJ} for an extensive exercise comparing techniques across many data generating processes). Finally, our approach is transparent to understand, easy to implement, and cheap to compute with existing statistical software commands.

\section{Relationship to Welfare Analysis\label{sec:MVPF}}

In this section, we discuss how supercomplier characteristics can be used to enrich welfare analyses. Specifically, documenting the differences between supercompliers and compliers informs how social welfare would change if a social planner put different ``weights” on individuals of different circumstances. More concretely, we describe how social weights are (and are not) used in the widely adopted framework of \citet{HendrenSprungKeyser2020}, and elaborate on how to appropriately weight treatment effects in this framework using supercomplier characteristics. 

\citet{HendrenSprungKeyser2020} advocate for the systematic reporting of the marginal value of public funds, or MVPF, to facilitate comparisons of public policies. The starting point for the MVPF is that the government is interested in the impact of a policy on social welfare, defined as the weighted sum of individual utilities:
\[
W=\sum_{i}\eta_{i}U_{i}.
\]
$U_{i}$ is expressed as a money-metric, and $\eta_{i}$ is the social welfare impact of transferring \$1 to individual $i$.\footnote{Any utility function can be normalized to a money-metric
by dividing by the marginal utility of income.} A small policy change, denoted by $dp$, impacts social welfare by:
\begin{equation}
\frac{dW}{dp}=\sum_{i}\eta_{i}\frac{dU_{i}}{dp}\equiv\sum_{i}\eta_{i}WTP_{i},\label{eq:weighted_wtp}
\end{equation}
where $WTP_{i}$ is interpreted as individual $i$'s ``willingness to pay'' (WTP) for the policy. Equation (\ref{eq:weighted_wtp}) encapsulates the benefits of the policy change. The cost of the policy equals its impact on the government's budget, denoted by $G$. \citet{HendrenSprungKeyser2020} define the MVPF formally as
\[
MVPF=\frac{\sum_{i}WTP_{i}}{G},
\]or, the ratio of the (unweighted) societal willingness to pay to the government net costs. Note that while the benefits of a policy change should incorporate social weights, the formal definition of the MVPF leaves the weights out, presumably because it is difficult to practically implement. \citet{HendrenSprungKeyser2020} suggest, therefore, to use MVPFs to compare policies that benefit similar beneficiaries, so as to ``reduce the role of social preferences in driving conclusions.'' As shown below, we can use supercomplier characteristics to incorporate weights into the MVPF to reflect a social planner's redistributional preferences, so long as the weights are functions of the characteristics reported.

The MVPF of a policy is calculated by plugging in estimates of its effect on the government budget (denominator) and, in many cases, the willingness to pay (numerator).n a typical application, the policy's effect on beneficiaries' (and possibly their children's) earnings are translated into an estimate of changes in tax revenue collected. The denominator may also incorporate spillover effects on government spending, such as changes in healthcare utilization among Medicare and Medicaid beneficiaries, changes in public school enrollment, or changes in individual contact with the criminal justice system. The willingness to pay is the dollar value that is transferred to individuals (before behavioral responses, per the envelope theorem) for policies that include a transfer or tax. However, if a policy affects later-life outcomes (e.g., additional education or improved health), causal estimates of these effects could better capture social benefits and therefore enter the numerator. 

The use of causal estimates as ``plug-in'' components of an MVPF calculation means that it corresponds to the subpopulation for which the causal estimates are identified. In particular, when using estimates from a randomized controlled trial with imperfect compliance as inputs, the resulting MVPF is the value of the policy for the (treatment) complier group.

A social planner may desire to place different weights on individuals' willingness to pay, even within the complier group. Suppose that the WTP is measured by the effect of the policy on an outcome $Y$, so that $WTP_{i}=Y_{1i}-Y_{0i}$. When $Y$ is binary, it is easy to see that the weighted WTP for the complier group is \begin{equation}
E[\eta_{i}(Y_{1i}-Y_{0i})|\text{complier}]=E[\eta_{i}|\text{supercomplier}]\text{LATE}_{Y}\label{eq:sc_wt}
\end{equation}
In other words, the weighted WTP is simply the (unweighted) local average treatment effect on the outcome of interest ($\text{LATE}_Y$), multiplied by the average social welfare weight of the supercompliers. We show in Appendix \ref{sec_app:sc_weight_cts} that this logic extends to the case where $Y$ is not binary. An important implication of this result is that if the social weights $\eta_{i}$ are a linear function of observable characteristics $X_{i}$, one can calculate the appropriately weighted WTP (and corresponding MVPF) from reported supercomplier characteristics. Linearity need not be restrictive in practice. In our empirical applications below, for example, baseline family income was recorded in predefined bins. By including an indicator variable for each bin (i.e., a fully saturated set of dummies), we effectively allow the welfare weights to vary flexibly across income categories, and the specification remains linear. 

While it is possible to simply estimate the weighted willingness to pay directly by estimating the treatment effect on weighted outcomes, the formulation in (\ref{eq:sc_wt}) shows that it can be done without microdata. If supercomplier characteristics are documented along with the LATE, a reader can construct a customized weighted WTP and MVPF themselves, to the extent that the desired social weights are a linear function of those reported characteristics. A major advantage of this approach is that it allows for readers to compute the MVPF using different social weights, which may not necessarily be the same as the weights chosen by the researchers who estimate the relevant treatment effect.

\section{Empirical Applications\label{sec:empirical}}

To demonstrate the utility our tools for characterizing supercompliers, we consider experimental evaluations of the Job Corps and Job Training Partnership Act programs.

\subsection{Job Corps}

Job Corps is a federally sponsored job training program for disadvantaged youths in the United States. The program offers a suite of services, including academic education, vocational training, and job search assistance. In the 1990s, Department of Labor sponsored the National Job Corps Study, where eligible Job Corps applicants between ages 16 and 24 were randomly assigned access to Job Corps. Researchers have used this experiment to study the effects of Job Corps on various outcomes (\citealt{JCImpacts2001}).

Consistent with Job Corps's mission, there were large impacts on educational attainment during the 48-month follow-up period in the public use data (\citealt{Schochet2008Data}). Members of the treatment group were 22.3 percentage points more likely to obtain a vocational certificate (37.5 percent compared to 15.2 percent; $t$-stat: 27.1; $N$: 11,151). Among students without a high school credential at random assignment, members of the treatment group were 15.0 percentage points more likely to obtain a GED (41.6 vs. 26.6 percent; $t$-stat: 14.3; $N$: 8,579). 
Job Corps also improved labor market outcomes, though the effects are more moderate. 
Specifically, the intent-to-treat effects on employment and earnings in the 16th quarter after random assignment---the final quarter of study by \citet{JCImpacts2001}---were 2.4 percentage points (71.1 vs. 68.7 percent; $t$-stat: 2.59; $N$: 10,872) and \$266, or 11 percent of the control mean, ($t$-stat: 4.76; $N$: 10,872), respectively. 

In Table \ref{tab:jc_edu}, we present supercomplier and complier characteristics with respect to educational attainment, as well as differences in characteristics between supercompliers and both compliers and the full experimental population (standard errors for the differences are estimated via stacked regressions). In Panel A, we examine characteristics for those attaining a GED, restricting to the sample without high school credentials prior to random assignment. In Panel B, we examine characteristics for those attaining a vocational credential. In both panels, we define (treatment) compliers as those who ultimately enrolled in Job Corps. They made up 73 percent and 72 percent of the respective samples.  
Panel A shows that supercompliers with respect to GED attainment were more likely to be female and white, and also to have had employment prior to random assignment, all relative to both the full population and compliers. Although not statistically significant at the 5 percent level, supercompliers also appeared to be slightly more advantaged with respect to family income. Similar patterns hold for supercompliers with respect to vocational certificate attainment (Panel B), and here we also find significant differences in arrest history, where supercompliers were less likely to have had a prior arrest. 

In Table \ref{tab:jc_empearn}, we present supercomplier characteristics with respect to a labor market outcome. Because of the relatively weak intent-to-treat effect on employment reported above, we focus on earnings in the 16th quarter after random assignment. The patterns in Table \ref{tab:jc_empearn} share many similarities with Table \ref{tab:jc_edu}: supercompliers were more likely to be white, older, less likely to have been arrested, and more likely to have had an employment history. But the magnitude differences are much more pronounced for two of these characteristics. Compared to Table \ref{tab:jc_edu}, supercompliers were substantially more likely to be white and age 20 or older. While not statistically significant at the 5 percent level, point estimates suggest that supercompliers were also more likely to come from higher-income families. Finally, for all outcomes in Tables \ref{tab:jc_edu} and \ref{tab:jc_empearn}, our tests based on Propositions \ref{prop:sharp} and \ref{prop:generalized_test} cannot reject our identifying assumptions at the 5 percent level, indicating their plausibility.

Overall, our labor market findings combined with our results on educational attainment suggest that participants who benefited most from Job Corps tended to be those who were less marginalized upon application: they were less likely to have had an arrest history, more likely to have been white, more likely to have had an employment history, and tended to be from families with higher income. We return to this point and its normative implications at the end of Section \ref{subsec:JTPA}.

\subsection{Job Training Partnership Act}
\label{subsec:JTPA}
The National Job Training Partnership Act (JTPA) Study was a randomized controlled trial evaluating job training programs in 16 sites across the US in the late 1980s. The JTPA programs primarily target economically disadvantaged adults and out-of-school youths. Eligible applicants to each site's JTPA program were randomly assigned to a control group or a treatment group, which was offered one of three service types: classroom training, subsidized on-the-job training, or other services. Following \citet{Bloom_etal1997}, we focus on earnings in the 30-month follow-up period as the main outcome of interest. \citet{HendrenSprungKeyser2020} also rely on this outcome for the MVPF calculation of JTPA. 

While \citet{Bloom_etal1997} estimated program effects for four distinct groups---adult men, adult women, male youths, and female youths---we restrict our analysis to adult women.\footnote{The treatment take-up effects were around 60 percent for all four groups.} Among the four groups, adult women saw the largest impact of the intervention with an intent-to-treat effect of \$1,190, or 10 percent of the control mean ($t$-stat: 3.63; $N$: 6,102). Adult men saw an intent-to-treat effect of \$1,051, or 6 percent of the control mean ($t$-stat: 2.01; $N$: 5,102). The intent-to-treat effects for the youth groups were insignificant at the 5 percent level. Since the intent-to-treat effect is the supercomplier ``first stage,'' it is only strong for adult women by the conventional rule of thumb of $F$-stat $>10$.\footnote{Recent studies by \citet{Leeetal2022} and \citet{Angristandkolesar2024} reexamine the reliability of inference procedures for just identified single-instrumental-variable models. We first note that the ``first-stage'' $F$-stats (square of the $t$-stats of the intent-to-treat effects) for the two panels in Table \ref{tab:jc_edu} are sufficiently strong (above the threshold of 142.6 according to \citealt{Leeetal2022}) to dispel inference concerns. In addition to the statistics reported in Tables \ref{tab:jc_empearn} and \ref{tab:jtpa}, we have also examined the Anderson-Rubin (AR) confidence intervals for supercomplier characteristics: the 95 percent AR confidence intervals are 9 to 13 percent (17 to 25 percent) wider for the characteristics reported in Table \ref{tab:jc_empearn} (Table \ref{tab:jtpa}) than their conventional counterparts. However, we can show that the statistical significance (at the 5 percent level) of the differences for white and older youths in Table \ref{tab:jc_empearn} is unchanged with AR intervals by applying a conservative test that results from the Bonferroni inequality. \citet{Angristandkolesar2024} advocate for screening for the sign of the ``first stage'' coefficient, and all of our intent-to-treat effects have the expected signs.}

Following the same format as Tables \ref{tab:jc_edu} and \ref{tab:jc_empearn}, Table \ref{tab:jtpa} reports the characteristics of the adult women in the study. We find that compliers were similar to the experimental population, but that supercompliers were somewhat different. With the caveat that standard errors are large and that none of the differences are statistically significant at the 5 percent level, we see that supercompliers were less attached to the labor force at baseline: they had lower prior earnings and weeks worked, were more likely to be on Aid to Families with Dependent Children (AFDC, i.e., cash assistance), and had lower family income.\footnote{Note that estimated characteristics shares of supercompliers are not guaranteed to lie between zero and one (in fact, neither are complier characteristics estimates). It is reassuring that none of our point estimates are negative or larger than one in Tables \ref{tab:jc_edu}-\ref{tab:jtpa}.} As with Job Corps, the test based on Proposition \ref{prop:generalized_test} also cannot reject our identifying assumptions in the JTPA data for the earnings outcome.  

We now turn to the normative implications of these results. Following the framework from Section \ref{sec:MVPF}, we estimate the JTPA program's socially weighted MVPF. \citet{HendrenSprungKeyser2020} estimate that the \emph{unweighted} MVPF of the JTPA program for adults (pooling men and women) is 1.38. This is calculated by assuming that participants value the program by their post-tax earnings impacts, net of AFDC benefits. Since the earnings impacts for women appear to be driven by less advantaged participants, and to the extent that one places a larger social weight on these participants, the weighted MVPF will be higher.\footnote{While the WTP calculated by \citet{HendrenSprungKeyser2020} contains earnings and AFDC impacts for both men and women, we only weight the earnings impacts for women. As mentioned above, the reduced form earnings effect for men is not strong enough to compute supercomplier weights. The same is true of AFDC impacts for both men and women. Furthermore, JTPA impacts on AFDC contribute minimally to the MVPF calculation: without accounting for it, the unweighted MVPF is 1.35 (vs. 1.38).}

To compute the weighted WTP (the numerator of the weighted MVPF), we first estimate the average social weight of the supercompliers. We assume a textbook social welfare function $\Psi(u)=\frac{u^{1-\phi}}{1-\phi}$ \citep{Salanie2011}, where $\phi$ denotes the extent to which the social planner desires redistribution and $u$ is income. A value of $\phi=0$ means that a dollar transfer has the same impact on social welfare regardless of income, while larger values of $\phi$ imply greater preference for redistribution. For a given $\phi$, the estimation of the average supercomplier social weight proceeds in three steps:
\begin{enumerate}
    \item Calculate the marginal social welfare of transferring \$1 to someone at the midpoint of each of the five income bins. That is, calculate $u^{-\phi}$ by setting $u$ equal to each midpoint. We truncate the top income bin (greater than \$12K) to (\$12K,\$15K) for the purpose of calculating marginal utilities.\footnote{This truncation artificially lowers the assumed income for individuals in the top bin, thereby inflating their marginal utility, resulting in a higher social weight for this income bin, and biasing against our results.}
    \item Calculate the social weights by normalizing the marginal welfare so that they average to one across the five income bins. For example, when $\phi=0.5$, the weights are 1.83, 1.05, 0.82, 0.69, and 0.61, respectively.
    \item Multiply each weight with the corresponding supercomplier share from Table \ref{tab:jtpa} and sum the products to arrive the average supercomplier social weight. When $\phi=0.5$, the average weight estimate is 1.29. 
\end{enumerate}
After these three steps, we multiply the average supercomplier weight with the reported LATE as per equation (\ref{eq:sc_wt}), and plug the result into the companion program by \citet{HendrenSprungKeyser2020} to arrive at the weighted MVPF. We find that when $\phi=0.5$, the weighted MVPF is 1.63, while $\phi=1$ implies a weighted MVPF of 1.97. 

As discussed above, this MVPF exercise depends on the normalization of the weights. Had we included additional higher income bins, supercompliers would receive an even greater weight, driving up the MVPF. The bottom line is that, whatever the choice of social welfare weights, it is possible to compute the weighted MVPF using only the reported supercomplier characteristics, provided that the weights are a linear function of those characteristics.

Finally, we note the contrast between the results for the JTPA and Job Corps studies. The earnings supercompliers (among women) in the JTPA study are less advantaged compared to the experimental population while the opposite is true in the Job Corps study. Although we do not have the data to estimate a weighted MVPF for Job Corps that directly corresponds to \citet{HendrenSprungKeyser2020}, our results based on publicly available survey data indicate that incorporating social welfare weights would decrease the MVPF for Job Corps youths by up to 43 percent depending on the value of $\phi$.\footnote{\citet{HendrenSprungKeyser2020} estimate an MVPF of 0.18 for Job Corps based on 20-year follow-up impacts from \citet{Schochet2018}, which come from tax data that are not publicly available. The impacts on earnings from the tax data are more muted even in the short run, and one explanation by \citet{Schochetetal2008} is that survey data includes informal earnings not captured by tax data.} In contrast, the MVPF with social welfare weights increases by up to 43 percent for the JTPA adults as shown above.

\section{Conclusion\label{sec:conclusion}}

In this paper, we develop methods to characterize the supercomplier subpopulation in a canonical instrumental variable framework with a binary instrument and treatment. We show that under the standard LATE assumptions plus outcome monotonicity, we can identify the characteristics of supercompliers. Because the plausibility of our identifying assumptions may depend on context, we develop sharp joint tests of their validity. In the presence of outcome monotonicity violations, we show that the average characteristics we identify are a linear combination of the characteristics of those benefiting from and those harmed by treatment receipt, implying that the bias will be small when few individuals are harmed by treatment receipt. Our identification results lead to natural estimators, which can be easily implemented using instrumental variable regressions via existing Stata commands. Finally, we show that supercomplier characteristics can be used to incorporate social weights in a welfare analysis of the marginal value of public funds. 

We illustrate the utility of our tools using data from two job-training experiments, the National Job Corps Study and the National Job Training Partnership Act Study. We find that participants whose labor market outcomes were improved by Job Corps had relatively higher baseline incomes, while the adult women who benefited from JTPA had relatively lower baseline incomes. To the extent that a social planner values redistribution, our results imply a lower MVPF for Job Corps and a higher MVPF for JTPA.

\newpage

\begin{singlespace}

\bibliographystyle{aea_fix_comma}
\bibliography{supercompliers.bib}

\end{singlespace}

\clearpage{}

\begin{table}[ht] 
  \centering
  \caption{Extended Principal Stratification}
    \begin{NiceTabular}{ccccccc}
    $G$-value & Treatment Type & Outcome Type & $D_0$   & $D_1$   & $Y_0$   & $Y_1$ \\
    \midrule
\textit{\textbf{aa}}    & \textbf{Always Taker} & \textbf{Always Taker} & \textbf{1}     & \textbf{1}     & \textbf{1}     & \textbf{1} \\
\textit{\textbf{an}}    & \textbf{Always Taker} & \textbf{Never Taker} & \textbf{1}     & \textbf{1}     & \textbf{0}     & \textbf{0} \\
\textit{\textbf{ac}}    & \textbf{Always Taker} & \textbf{Complier} & \textbf{1}     & \textbf{1}     & \textbf{0}     & \textbf{1} \\
  \emph{af}    & Always Taker & Defier & 1     & 1     & 1     & 0 \\
\textit{\textbf{na}}    & \textbf{Never Taker} & \textbf{Always Taker} & \textbf{0}     & \textbf{0}     & \textbf{1}     & \textbf{1} \\
\textit{\textbf{nn}}    & \textbf{Never Taker} & \textbf{Never Taker} & \textbf{0}     & \textbf{0}     & \textbf{0}     & \textbf{0} \\
\textit{\textbf{nc}}    & \textbf{Never Taker} & \textbf{Complier} & \textbf{0}     & \textbf{0}     & \textbf{0}     & \textbf{1} \\
    \emph{nf}    & Never Taker & Defier & 0     & 0     & 1     & 0 \\
\textit{\textbf{ca}}    & \textbf{Complier} & \textbf{Always Taker} & \textbf{0}     & \textbf{1}     & \textbf{1}     & \textbf{1} \\
\textit{\textbf{cn}}    & \textbf{Complier} & \textbf{Never Taker} & \textbf{0}     & \textbf{1}     & \textbf{0}     & \textbf{0} \\
\textit{\textbf{cc}}    & \textbf{Complier} & \textbf{Complier} & \textbf{0}     & \textbf{1}     & \textbf{0}     & \textbf{1} \\
    \emph{cf}    & Complier & Defier & 0     & 1     & 1     & 0 \\
    \emph{fa}    & Defier & Always Taker & 1     & 0     & 1     & 1 \\
    \emph{fn}    & Defier & Never Taker & 1     & 0     & 0     & 0 \\
    \emph{fc}    & Defier & Complier & 1     & 0     & 0     & 1 \\
    \emph{ff}    & Defier & Defier & 1     & 0     & 1     & 0 \\
    \midrule
    \end{NiceTabular}%
     \label{tab:sub}
\end{table}%

\newpage{}
\begin{table}[t]
\caption{Job Corps Supercomplier Characteristics for Education Outcomes}
\label{tab:jc_edu}
\vspace{-20bp}
\includegraphics[viewport=4bp 105bp 594bp 756bp,scale=0.9]{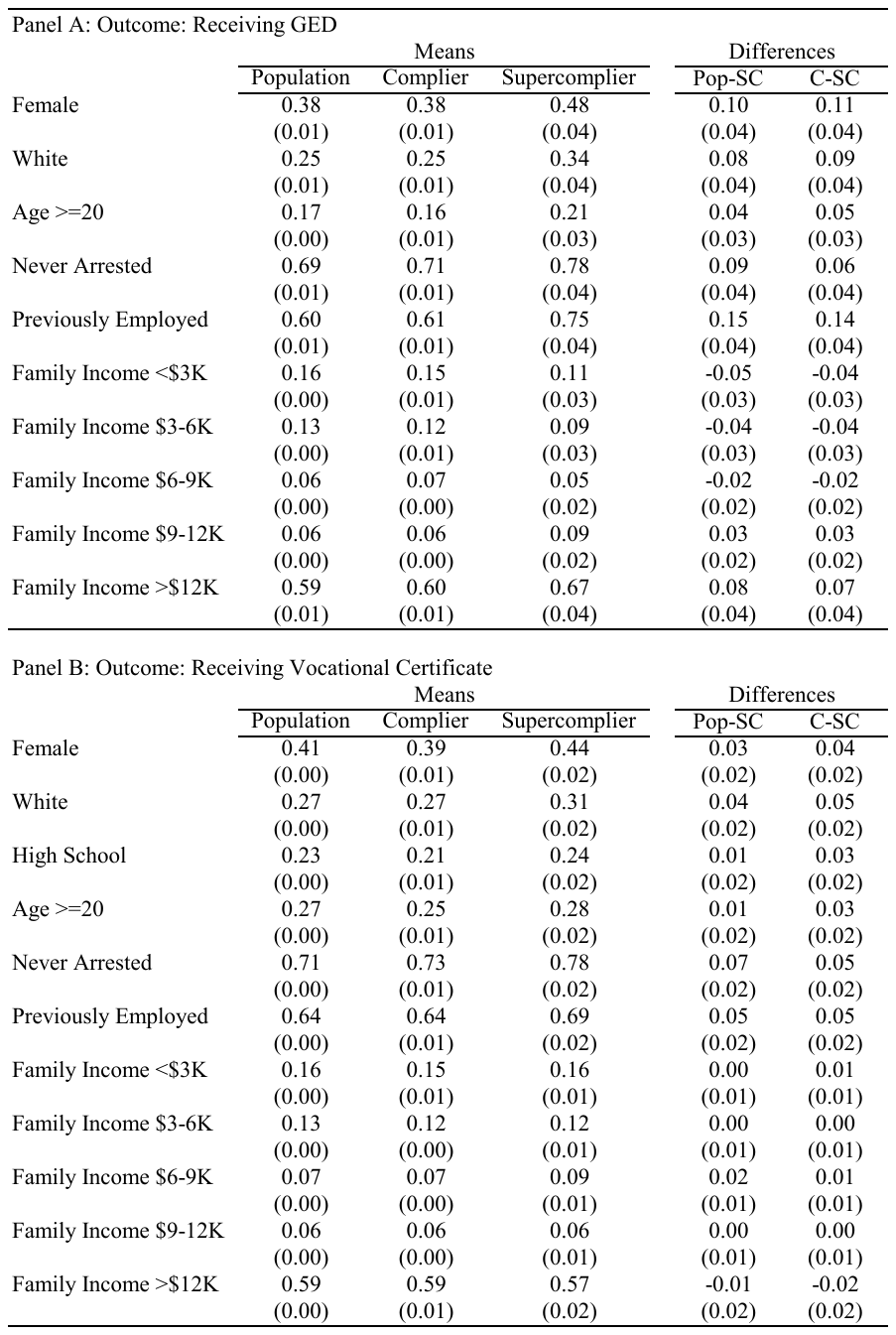}
\begin{centering}
\begin{minipage}[t]{0.82\columnwidth}%
\small{Notes: This table reports the population (Pop), complier (C), and supercomplier (SC) averages in baseline characteristics as well as the Pop-SC and C-SC differences. As with \citet{JCImpacts2001} and \citet{Schochetetal2008}, we use weights included in the public use data to adjust for sample and survey designs. Standard errors are in parentheses.}
\end{minipage}
\par\end{centering}
\end{table}

\newpage{}

\begin{table}[t]
\caption{Job Corps Supercomplier Characteristics for Earnings}
\label{tab:jc_empearn}
\vspace{-20bp}
\includegraphics[viewport=4bp 416bp 594bp 756bp,scale=0.90]{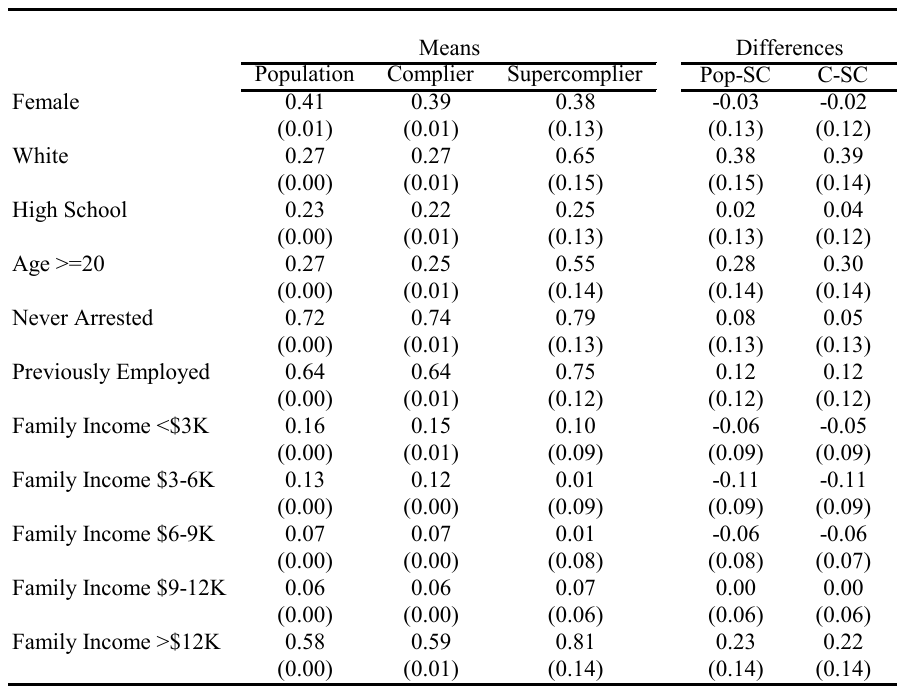}
\begin{centering}
\begin{minipage}[t]{0.82\columnwidth}%
\small{Notes: This table reports the population (Pop), complier (C), and supercomplier (SC) averages in baseline characteristics as well as the Pop-SC and C-SC differences for the outcome of earnings during the 16th quarter after random assignment (quarter 16 is the final quarter of study for the initial Job Corps assessment by \citealt{JCImpacts2001}). As with \citet{JCImpacts2001} and \citet{Schochetetal2008}, we use weights included in the public use data to adjust for sample and survey designs. Standard errors are in parentheses.}
\end{minipage}
\par\end{centering}
\end{table}

\newpage{}

\begin{table}[t]
\caption{National JTPA Study Supercomplier Characteristics for Earnings}
\label{tab:jtpa}
\vspace{-20bp}
\includegraphics[viewport=11bp 310bp 594bp 756bp,scale=0.9]{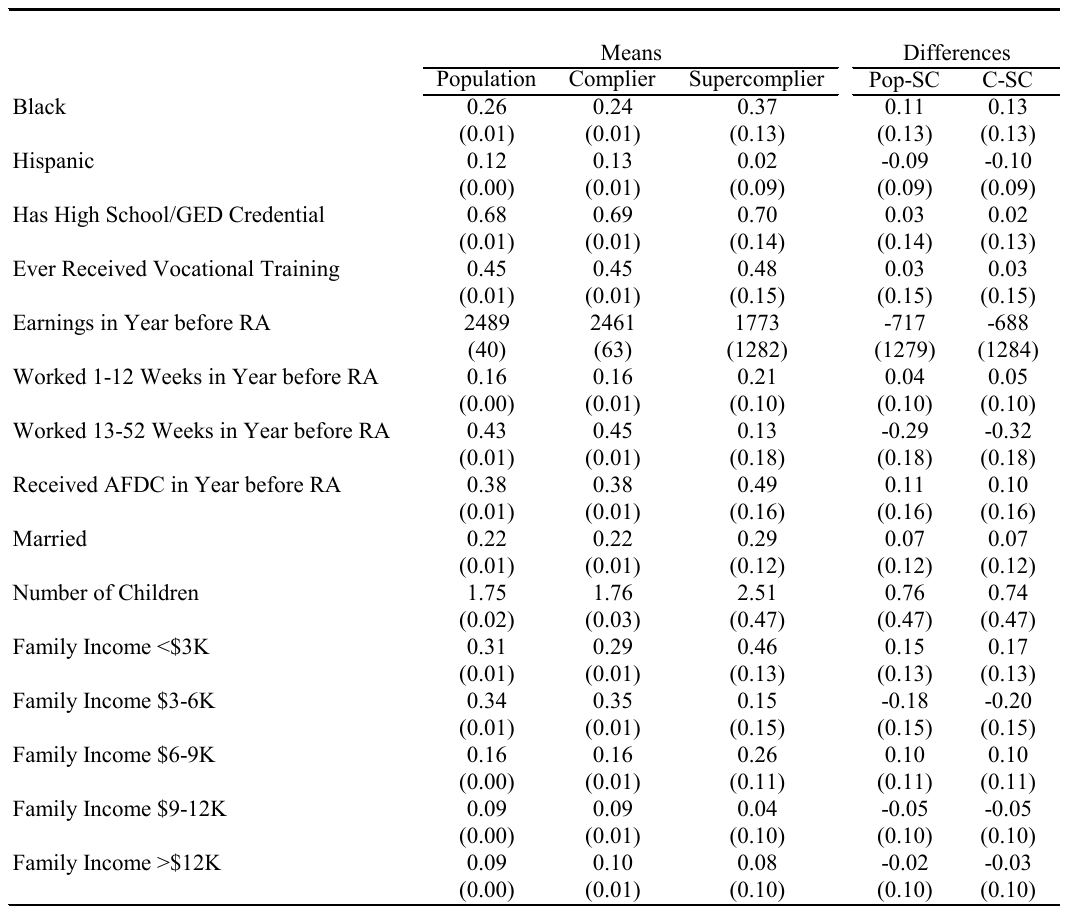}
\begin{centering}
\begin{minipage}[t]{0.95\columnwidth}%
\small{Notes: This table reports the population (Pop), complier (C), and supercomplier (SC) averages in baseline characteristics as well as the Pop-SC and C-SC differences. Our regressions control for the same covariates as in the impact analysis by \citet{Bloom_etal1997}. Standard errors are in parentheses. ``RA'' is short for random assignment. Data source: \citet{Orretal2003Data}.}
\end{minipage}
\par\end{centering}
\end{table}

\clearpage{}

\newpage{}

\pagenumbering{arabic}

\renewcommand{\theequation}{A\arabic{equation}}

\setcounter{equation}{0}

\setcounter{footnote}{0}

\setcounter{figure}{0} \renewcommand{\thefigure}{A.\arabic{figure}} 
\setcounter{table}{0} \renewcommand{\thetable}{A.\arabic{table}} 

\appendix
\begin{center}
\textbf{\Large{}Appendix (For Online Publication Only)}{\Large\par}
\par\end{center}
\section{Additional Theoretical Results}

\subsection{Proofs of Lemma \ref{lemma:decomp} and Propositions \ref{prop:SC_Char}-\ref{prop:No_OM} \label{subsec:proofs}}

{\bf Proof of Lemma \ref{lemma:decomp}}: We start from the right hand side of the equation
\begin{align*}
    &E[Y|Z=1]-E[Y|Z=0]\\
    =&\Pr(Y=1|Z=1)-\Pr(Y=1|Z=0)\\
    =&\Pr(G=aa,ac,na,ca,cc|Z=1)-\Pr(G=aa,ac,na,ca|Z=0)\\
    =&\Pr(G=cc).
\end{align*}
The first equality follows from $Y$ being binary. The second equality follows from Assumptions \ref{A:IV}.2, \ref{A:IV}.3, and \ref{A:OM}.1. The last equality follows from Assumption \ref{A:IV}.1 and the fact that the groups in Table \ref{tab:sub} are mutually exclusive. \emph{QED.}

To prove Proposition \ref{prop:SC_Char}, we first state a lemma that provides the analog of Theorem 3.1 of \citet{Abadie2003} under our Assumption \ref{A:IV}. 

\begin{lemma}
\label{lemma:Abadie}
Let $g(\cdot)$ be any function of $(Y,D,X)$ such that $E[|g(Y,D,X)|]<\infty$. Under Assumption 1, provided that $X\indep Z$, and with $\kappa$, $\kappa_0$, and $\kappa_1$ defined in Proposition \ref{prop:SC_Char},

(a) $E[g(Y,D,X)|D_{1}>D_{0}]=\frac{1}{\Pr(D_{1}>D_{0})}E[\kappa g(Y,D,X)]$. Also,

(b) $E[g(Y_{0},X)|D_{1}>D_{0}]=\frac{1}{\Pr(D_{1}>D_{0})}E[\kappa_{0}g(Y,X)]$, and

(c) $E[g(Y_{1},X)|D_{1}>D_{0}]=\frac{1}{\Pr(D_{1}>D_{0})}E[\kappa_{1}g(Y,X)].$
\end{lemma}

We omit the proof of Lemma \ref{lemma:Abadie} as it follows the same line of reasoning as the proof of Theorem 3.1 in \citet{Abadie2003}. However, Lemma \ref{lemma:Abadie} is not a special case of Theorem 3.1 of \citet{Abadie2003}. \citet{Abadie2003} assumes conditional independence, and only the covariates in the conditioning set appear in his Theorem 3.1. We assume unconditional independence, which means our conditioning set is empty, but we still have covariates in Lemma \ref{lemma:Abadie}. We can generalize Lemma \ref{lemma:Abadie} to nest both our unconditional case and \citet{Abadie2003} by allowing for an arbitrary conditioning set and additional covariates not in the conditioning set.

{\bf Proof of Proposition \ref{prop:SC_Char}}: To prove the identification result in equation (\ref{eq:sc_char_id_weight}), we note that under Assumption \ref{A:OM}, the treatment complier population consists of three groups: $G={ca,cn,cc}$. Thus,
\begin{align}
&E[h(X)|G=cc]\Pr(G=cc|D_{1}>D_{0})\nonumber\\
=&E[h(X)|D_{1}>D_{0}]-E[h(X)|G=ca]\Pr(G=ca|D_{1}>D_{0}) \nonumber \\
& \; \; \; - E[h(X)|G=cn]\Pr(G=cn|D_{1}>D_{0}).\label{eq:SC_expand}  
\end{align}
We examine each of the three terms on the right-hand side of equation (\ref{eq:SC_expand}). The first term is simply the average $h(X)$ for the treatment compliers. Per Lemma \ref{lemma:Abadie}(a),
\begin{equation*}
E[h(X) | D_1 > D_0] = \frac{1}{\Pr(D_1 > D_0)} E[\kappa h(X)].
\end{equation*}
For the second term,
\begin{align*}
& E[h(X) | G=ca] \Pr(G=ca | D_1 > D_0) \\
=&E\left[h(X)|D_{1}>D_{0},Y_{0}=Y_{1}=1\right]\Pr(Y_{0}=Y_{1}=1|D_{1}>D_{0})\\
\overset{\text{(i)}}{=}&E\left[h(X)|D_{1}>D_{0},Y_{0}=1\right]\Pr(Y_{0}=1|D_{1}>D_{0})\\
=&E\left[h(X)Y_{0}|D_{1}>D_{0},Y_{0}=1\right]\Pr(Y_{0}=1|D_{1}>D_{0})\\
=&E\left[h(X)Y_{0}|D_{1}>D_{0}\right]\\
\overset{\text{(ii)}}{=}&\frac{1}{\Pr(D_{1}>D_{0})}E[\kappa_{0}Yh(X)],
\end{align*}
where equality (i) follows from Assumption \ref{A:OM}.1 and (ii) from Lemma \ref{lemma:Abadie}(b). 

Analogously, for the third term, we can show that 
\begin{equation*}
E[h(X)|G=cn]\Pr(G=cn|D_{1}>D_{0})=\frac{1}{\Pr(D_{1}>D_{0})}E[\kappa_{1}(1-Y)h(X)].  
\end{equation*}
Plugging these results back into equation (\ref{eq:SC_expand}), we have 
\begin{equation*}
E[h(X)|G=cc]=\frac{E[\pi h(X)]}{\Pr(G=cc|D_{1}>D_{0})\Pr(D_{1}>D_{0})}=\frac{E[\pi h(X)]}{\Pr(G=cc)}. 
\end{equation*}
Because Lemma \ref{lemma:decomp} implies the denominator in the equation above to be the reduced form, $RF$, we have the desired result.

For equation (2), we note that it is a special case of Proposition \ref{prop:non_binary_Y}, which we prove next. Specifically, when $Y$ is binary, we can plug $Y_1-Y_0=1$ into equation (\ref{eq:char_non_binary_Y}) and obtain the desired result. \emph{QED.}

{\bf Proof of Proposition \ref{prop:non_binary_Y}}: For equation (\ref{eq:share_non_binary_Y}), 
\begin{align*}
    &E[Y|Z=1]-E[Y|Z=0]\\
    =&E[Y_{1}-Y_{0}|D_{1}>D_{0}]\Pr(D_{1}>D_{0})\\
    =&E[Y_{1}-Y_{0}|D_{1}>D_{0},Y_{1}>Y_{0}]\Pr(Y_{1}>Y_{0}|D_{1}>D_{0})\Pr(D_{1}>D_{0})\\
    =&\Pr(G=cc)E[Y_{1}-Y_{0}|G=cc].
\end{align*}
For equation (\ref{eq:char_non_binary_Y}), the first term in the numerator of the Wald estimand is
\begin{align*}
&E[h(X)Y|Z=1]\\
=&E[h(X)Y|Z=1,(D_1=D_0 \text{ or } Y_1=Y_0)]\Pr(D_1=D_0 \text{ or } Y_1=Y_0|Z=1)+\\
&E[h(X)Y|Z=1,G=cc]\Pr(G=cc|Z=1).
\end{align*}
We can similarly expand the second term in the numerator. The difference between the two terms is nonzero only for the supercomplier group. Therefore,
\begin{align*}
&E[h(X)Y|Z=1]-E[h(X)Y|Z=0]\\
=&E[h(X)Y|Z=1,G=cc]\Pr(G=cc|Z=1)-\\
&E[h(X)Y|Z=0,G=cc]\Pr(G=cc|Z=0)\\
=&E[h(X)(Y_1-Y_0)|G=cc]\Pr(G=cc).
\end{align*}
Plugging the left hand side of equation (\ref{eq:share_non_binary_Y}) into the denominator of the Wald estimand, we have the desired result. \emph{QED}.

{\bf Proof of Proposition \ref{prop:sharp}}: Since Proposition \ref{prop:sharp} is a special case of Proposition \ref{prop:generalized_test} in Appendix \ref{subsec:generalized_test} below, we omit its proof here. Interested readers can still find a proof specifically for the simpler Proposition \ref{prop:sharp} in a previous version of our paper \citet{Comeyetal2023}. 

{\bf Proof of Proposition \ref{prop:No_OM}}: The numerator of the estimand is
\footnotesize
\begin{align*}
&E[h(X)Y|Z=1]-E[h(X)Y|Z=0]\\
=&E[h(X)Y_{1}|Z=1,D_{0}=1]\Pr(D_{0}=1|Z=1)+E[h(X)Y_{0}|Z=1,D_{1}=0]\Pr(D_{1}=0|Z=1)+\\
&E[h(X)Y_{1}|Z=1,D_{1}>D_{0}]\Pr(D_{1}>D_{0}|Z=1)-E[h(X)Y_{1}|Z=0,D_{0}=1]\Pr(D_{0}=1|Z=0)-\\
&E[h(X)Y_{0}|Z=0,D_{1}=0]\Pr(D_{1}=0|Z=0)-E[h(X)Y_{0}|Z=0,D_{1}>D_{0}]\Pr(D_{1}>D_{0}|Z=0)\\
=&E[h(X)Y_{1}|D_{1}>D_{0}]\Pr(D_{1}>D_{0})-E[h(X)Y_{0}|D_{1}>D_{0}]\Pr(D_{1}>D_{0})\\
=&E[h(X)|D_{1}>D_{0},Y_{1}=1,Y_{0}=1]\Pr(Y_{1}=1,Y_{0}=1|D_{1}>D_{0})\Pr(D_{1}>D_{0})+\\
&E[h(X)|D_{1}>D_{0},Y_{1}=1,Y_{0}=0]\Pr(Y_{1}=1,Y_{0}=0|D_{1}>D_{0})\Pr(D_{1}>D_{0})-\\
&E[h(X)|D_{1}>D_{0},Y_{1}=1,Y_{0}=1]\Pr(Y_{1}=1,Y_{0}=1|D_{1}>D_{0})\Pr(D_{1}>D_{0})-\\
&E[h(X)|D_{1}>D_{0},Y_{1}=0,Y_{0}=1]\Pr(Y_{1}=0,Y_{0}=1|D_{1}>D_{0})\Pr(D_{1}>D_{0})\\
=&E[h(X)|G=cc]\Pr(G=cc)-E[h(X)|G=cf]\Pr(G=cf).
\end{align*}
\normalsize
Similarly, we can show that the denominator is
\footnotesize
\begin{equation*}
E[Y|Z=1]-E[Y|Z=0]=\Pr(Y_{1}>Y_{0},D_{1}>D_{0})-\Pr(Y_{1}<Y_{0},D_{1}>D_{0}).
\end{equation*}
\normalsize
Combining the numerator and denominator expressions and rearranging, we have our desired result. \emph{QED.}

\subsection{Shares and Characteristics of Other Groups\label{subsec_app:othergroups}}

Lemma \ref{lemma:decomp} and Proposition \ref{prop:SC_Char} show that we can identify the share and characteristics of the supercompliers. The proposition below establishes identification of the shares and characteristics of the other two groups within the stratum of compliers, $G=ca,cn$.
\begin{proposition}
\label{prop:other_groups}
Under Assumptions \ref{A:IV} and \ref{A:OM} and provided that $X \indep Z$, the shares of the two groups are
\begin{align*}
\Pr(G=ca) & =E[\kappa_{0}Y]=E[(1-D)Y|Z=0]-E[(1-D)Y|Z=1]\\
\Pr(G=cn) & =E[\kappa_{1}(1-Y)]=E[D(1-Y)|Z=1]-E[D(1-Y)|Z=0],
\end{align*}
and their characteristics can be identified as
\begin{align*}
E[h(X)|G=ca] & =\frac{E[\kappa_{0}Yh(X)]}{E[\kappa_{0}Y]}
=\frac{E[(1-D)Yh(X)|Z=1]-E[(1-D)Yh(X)|Z=0]}{E[(1-D)Y|Z=1]-E[(1-D)Y|Z=0]}\\
E[h(X)|G=cn] & =\frac{E[\kappa_{1}(1-Y)h(X)]}{E[\kappa_{1}(1-Y)]}=\frac{E[D(1-Y)h(X)|Z=1]-E[D(1-Y)h(X)|Z=0]}{E[D(1-Y)|Z=1]-E[D(1-Y)|Z=0]}.
\end{align*}
\end{proposition}
{\bf Proof of Proposition \ref{prop:other_groups}}: We only prove the identification results for $G=ca$, as the proof for $G=cn$ is analogous. 
For the share of the $G=ca$ group, first note that 
\begin{equation*}
    \Pr(G=ca)\equiv\Pr(Y_{0}=1,D_{1}>D_{0})=E[Y_{0}|D_{1}>D_{0}]\Pr(D_{1}>D_{0})=E[\kappa_{0}Y],
\end{equation*}
where the last equality follows from Lemma \ref{lemma:Abadie}(b). This proves the first part of the equality, and the second part can be easily proved by plugging in the definition of $\kappa_0$ and simplifying.
For the characteristics result, note that
\begin{equation*}
E[h(X)|G=ca]=E[h(X)|D_{1}>D_{0},Y_{0}=1]=\frac{E[Y_{0}h(X)|D_{1}>D_{0}]}{E[Y_{0}|D_{1}>D_{0}]}    
\end{equation*}
Since Lemma \ref{lemma:Abadie}(b) also implies
\begin{equation*}
E[Y_{0}h(X)|D_{1}>D_{0}]=\frac{1}{\Pr(D_{1}>D_{0})}E[\kappa_{0}Yh(X)],    
\end{equation*}
we have the first part of the identification result:
\begin{equation*}
    E[h(X)|G=ca]=\frac{E[\kappa_{0}Yh(X)]}{E[\kappa_{0}Y]}
\end{equation*}
We obtain the second part of the equality, i.e., identification by the Wald-type estimand, by plugging in the definition of $\kappa_0$ and simplifying. \emph{QED.}

Like Proposition \ref{prop:SC_Char}, the identification result in Proposition \ref{prop:other_groups} leads to natural estimators. We can estimate the population share of the $G=ca$ group by regressing $(1-D)Y$ on $(1-Z)$ and the population share of the $G=cn$ group by regressing $D(1-Y)$ on $Z$. With the Wald-type estimand for the characteristics, we can implement the corresponding 2SLS estimators like we do for the supercomplier characteristics. 

Finally, we note that it is easy to directly identify the observations assigned to treatment (control) as belonging to the $G=na$ ($G=an$) group, namely those with $D=0$ and $Y=1$ ($D=1$ and $Y=0$). Therefore, we can identify the shares and characteristics of these two groups by directly using these observations. However, we cannot identify the shares and characteristics of the four remaining groups in Table \ref{tab:sub}, $G=aa,ac,nc,nn$. To see this, suppose for the sake of contradiction that we can identify the share of the $aa$ group. Because we can identify the share of the treatment always takers, we will also be able to identify the share of the $ac$ group (by subtraction) and, consequently, the probability of being in the $ac$ group conditional on being a treatment always taker. However, the latter is simply the treatment effect for the treatment always takers:
\begin{equation*}
    \Pr(G=ac|D_{0}=D_{1}=1)=E[Y_{1}-Y_{0}|D_{0}=D_{1}=1],
\end{equation*}
which we know cannot be identified. This contradiction shows the non-identifiability of the $aa$ group share, and the proofs for the other groups are analogous. 

\subsection{Adding a Mediator\label{subsec_app:mediator_details}}

In this section, we provide details for Remark \ref{remark:mediator}. To incorporate a mediator $M$, we first expand on Assumptions \ref{A:IV} and \ref{A:OM}. 

\begin{assumption}[Mediator] \label{A:IV-Mediator}
\leavevmode
\begin{enumerate}[label={\arabic*.}, align=left]
\item Random Assignment: $(\{Y_{zdm}\}, \{M_{zd}\}, \{D_z\}) \indep Z$ for $z,d,m=0,1$ and \\
$0 < \Pr(Z=1) < 1$. 
\item Exclusion: $\Pr(M_{1d}=M_{0d})=1$ for $d\in\{0,1\}$ and \\ 
$\Pr(Y_{zdm} = Y_{z^{\prime}d^{\prime}m}) = 1 \text{ for } \{zd\} \neq \{z^{\prime}d^{\prime}\} \text{ and } m\in\{0,1\}.$
\item Monotonicity: $\Pr(D_1 \geq D_0) = 1$, $\Pr(M_1 \geq M_0) = 1$, and $\Pr(Y_1 \geq Y_0) = 1$.
\item Causal Chain: $\Pr(Y_1 > Y_0, M_1 > M_0, D_1 > D_0) > 0.$
\end{enumerate}
\end{assumption}

The inclusion of mediator $M$ expands the dimensionality of the principal stratum partitioning from Table \ref{tab:sub}. With the monotonicity conditions in Assumption \ref{A:IV-Mediator}.3, we now have 27 admissible sub-groups, as opposed to nine admissible sub-groups under Assumptions \ref{A:IV} and \ref{A:OM} alone. We denote each of the groups in the expanded stratification as $\tilde{G}=dmy$ for $d,m,y=a,c,n$.

\begin{proposition}
\label{prop:mediator}
Under Assumption \ref{A:IV-Mediator} and provided that $X \indep Z$, 
\begin{enumerate}[label=(\Alph*)]
\item The $\tilde{G}=ccc$ group (``super-duper complier'') share is
\begin{equation*}
    \Pr(\tilde{G}=ccc)=E[Y|Z=1]-E[Y|Z=0];
\end{equation*}
and its mean characteristics 
\begin{equation*}
E[h(X) | \tilde{G}=ccc]  = \frac{E[h(X)Y | Z = 1] - E[h(X)Y | Z=0] }{ E[Y | Z=1] - E[Y|Z=0] }.
\end{equation*}
\item The $\tilde{G}=cca$ group share is
\begin{equation*}
    \Pr(\tilde{G}=cca)=E [ (1-M) Y |Z=0] - E [(1-M) Y |Z=1],
\end{equation*}
and its mean characteristics
\begin{equation*}
    E[h(X)|\tilde{G}=cca]=\frac{E[(1-M)Yh(X)|Z=1]-E[(1-M)Yh(X)|Z=0]}{E[(1-M)Y|Z=1]-E[(1-M)Y|Z=0]}
\end{equation*}
\item The $\tilde{G}=ccn$ group share is
\begin{equation*}
    \Pr(\tilde{G}=ccn)=E [ M(1-Y) |Z=1] - E [ M(1-Y)  |Z=0],
\end{equation*}
and its mean characteristics
\begin{equation*}
   E[h(X)|\tilde{G}=ccn]=\frac{E[M(1-Y) h(X)|Z=1]-E[M(1-Y) h(X)|Z=0]}{E[M(1-Y) |Z=1]-E[M(1-Y) |Z=0]}.
\end{equation*}
\end{enumerate}
\end{proposition}
{\bf Proof of Proposition \ref{prop:mediator}}:

\begin{enumerate}[label=(\Alph*)]

\item 
The key insight for this part is that $\tilde{G}=ccc$ the only group whose $Y$ changes with $Z$. The formal proof follows the same logic as that in the proof of Proposition \ref{prop:non_binary_Y} and is therefore omitted.

\item 
For the share $\Pr(\tilde{G}=cca)$, note that
\begin{align}
    &E[(1-M)Y|Z=0]\nonumber\\
    =&\Pr(D=1,M=0,Y=1|Z=0)+\Pr(D=0,M=0,Y=1|Z=0)\nonumber\\
=&\Pr(\tilde{G}=ana)+\Pr(\tilde{G}=cca,cna,nca,nna).\label{eq:cca1}
\end{align}
Similarly, we have
\begin{equation}
    E[(1-M)Y|Z=1]=\Pr(\tilde{G}=ana,cna)+\Pr(\tilde{G}=nca,nna).\label{eq:cca2}
\end{equation}
Taking the difference of (\ref{eq:cca1}) and (\ref{eq:cca2}), we have the desired result.

For average characteristics, the first term in the numerator of the Wald estimand is
\begin{align}
    &E[(1-M)Yh(X)|Z=1]\nonumber\\
    =&E[D(1-M)Yh(X)|Z=1]+E[(1-D)(1-M)Yh(X)|Z=1]\nonumber\\
    =&E[h(X)|\tilde{G}=ana,cna]\Pr(\tilde{G}=ana,cna)+\nonumber\\
    &E[h(X)|\tilde{G}=nca,nna]\Pr(\tilde{G}=nca,nna)\nonumber\\
    =&\sum_{\tilde{g}=ana,cna,nca,nna}E[h(X)|\tilde{G}=\tilde{g}]\Pr(\tilde{G}=\tilde{g}).\label{eq:cca3}
\end{align}
Similarly, we can show that 
\begin{equation}
    E[(1-M)Yh(X)|Z=0]=\sum_{\tilde{g}=ana,cna,nca,nna,cca}E[h(X)|\tilde{G}=\tilde{g}]\Pr(\tilde{G}=\tilde{g}).\label{eq:cca4}
\end{equation}
We have the desired result after taking the difference of (\ref{eq:cca3}) and (\ref{eq:cca4}) and replacing the denominator of the Wald estimand by $\Pr(\tilde{G}=cca)$.

\item 

The proof proceeds analogously to part (B) by noticing:
\begin{align*}
    E[M(1-Y)|Z=1]=&\Pr(\tilde{G}=aan,acn,can,ccn)+\Pr(\tilde{G}=nan)\\
    E[M(1-Y)|Z=0]=&\Pr(\tilde{G}=aan,acn)+\Pr(\tilde{G}=nan,can).\text{ }QED.
\end{align*}
\end{enumerate}

\subsection{Estimator Details\label{subsec_app:estimator_details}}

We prove several claims about complier and supercomplier characteristics estimators from Section \ref{subsec:estimation}. First, we show that the \citet{AngristandPischke2009} plug-in estimator for complier characteristics and the analogous estimator for supercomplier characteristics have Wald-type representations. These plug-in estimators are defined, respectively, as:
\begin{align*}
    \tilde{\chi}_{c}&\equiv\frac{\sum_{i}\hat{\kappa}_{i}h(X_{i})}{\sum_{Z_{i}=1}D_{i}-\sum_{Z_{i}=0}D_{i}}\\
    \tilde{\chi}_{cc}&\equiv\frac{\sum_{i}\hat{\pi}_{i}h(X_{i})}{\sum_{Z_{i}=1}Y_{i}-\sum_{Z_{i}=0}Y_{i}},
\end{align*}
where $\hat{\kappa}_i$ and $\hat{\pi}_i$ are sample analogs of the $\kappa$ and $\pi$ weighting functions evaluated at the $D_i$, $Z_i$, and $Y_i$ for individual $i$.

Define $\hat{\chi}_{cc}$ as the Wald-type estimator associated with equation (\ref{eq:sc_char_id_wald}) and $\hat{\chi}_{c}$ as the analogous treatment complier characteristics estimator:
\begin{align*}
\hat{\chi}_{c}&\equiv\frac{\sum_{Z_{i}=1}h(X_{i})D_{i}-\sum_{Z_{i}=0}h(X_{i})D_{i}}{\sum_{Z_{i}=1}D_{i}-\sum_{Z_{i}=0}D_{i}}\\
\hat{\chi}_{cc}&\equiv\frac{\sum_{Z_{i}=1}h(X_{i})Y_{i}-\sum_{Z_{i}=0}h(X_{i})Y_{i}}{\sum_{Z_{i}=1}Y_{i}-\sum_{Z_{i}=0}Y_{i}}.
\end{align*}
Our second claim is that $\hat{\chi}_{c}$ is equivalent to the complier characteristics estimator used by studies such as \citet{FinkNoto2019}.\footnote{It is easy to show that $\hat{\chi}_{c}$ is also the sample analog of $E[\kappa_1h(X)]/FS$ where $FS$ stands for the population first stage.} Finally, we compare the asymptotic variances of the two candidate supercomplier estimators, $\tilde{\chi}_{cc}$ and $\hat{\chi}_{cc}$, and show that neither dominates the other (the same is true for the two complier characteristics estimators).

\begin{proposition}
\label{prop:plug-in-estimators}
Let $N_1$ and $N_0$ denote the number of sample units assigned to the treatment and control groups, respectively, and $\hat{\tau}\equiv N_{1}/N$ the proportion in the treatment group. We can rewrite the plug-in estimators as 
\begin{align*}
\tilde{\chi}_{c}&=\frac{\frac{1}{N_{1}}\sum_{Z_{i}=1}\{D_{i}-(1-\hat{\tau})\}h(X_{i})-\frac{1}{N_{0}}\sum_{Z_{i}=0}\{D_{i}-(1-\hat{\tau})\}h(X_{i})}{\frac{1}{N_{1}}\sum_{Z_{i}=1}\{D_{i}-(1-\hat{\tau})\}-\frac{1}{N_{0}}\sum_{Z_{i}=0}\{D_{i}-(1-\hat{\tau})\}}\\
\tilde{\chi}_{cc}&=\frac{\frac{1}{N_{1}}\sum_{Z_{i}=1}\{Y_{i}-(1-\hat{\tau})\}h(X_{i})-\frac{1}{N_{0}}\sum_{Z_{i}=0}\{Y_{i}-(1-\hat{\tau})\}h(X_{i})}{\frac{1}{N_{1}}\sum_{Z_{i}=1}\{Y_{i}-(1-\hat{\tau})\}-\frac{1}{N_{0}}\sum_{Z_{i}=0}\{Y_{i}-(1-\hat{\tau})\}}
\end{align*}
\end{proposition}
{\bf Proof of Proposition \ref{prop:plug-in-estimators}}: We prove the result for $\tilde{\chi}_{c}$, and the proof for $\tilde{\chi}_{cc}$ is analogous. We begin by plugging into the numerator of $\tilde{\chi_{c}}$ the estimator of $\kappa_i$, for which $\Pr(Z=1)$ is estimated with $\hat{\tau}$:
\begin{align*}
\frac{1}{N}\sum_{i}\hat{\kappa}_{i}h(X_{i})&=\frac{1}{N}\sum_{i}\left[h(X_{i})-\frac{h(X_{i})D_{i}(1-Z_{i})}{1-\hat{\tau}}-\frac{h(X_{i})(1-D_{i})Z_{i}}{\hat{\tau}}\right]\\
&=\frac{1}{N}\sum_{i}h(X_{i})-\frac{1}{N}\sum_{i}\frac{h(X_{i})Z_{i}}{\hat{\tau}}+\frac{1}{N}\sum_{i}\frac{h(X_{i})D_{i}Z_{i}}{\hat{\tau}}-\frac{1}{N}\sum_{i}\frac{h(X_{i})D_{i}(1-Z_{i})}{1-\hat{\tau}}.
\end{align*}
For the last three terms,
\begin{align*}
    -\frac{1}{N}\sum_{i}\frac{h(X_{i})Z_{i}}{\hat{\tau}}&=-\frac{1}{N_{1}}\sum_{Z_{i}=1}h(X_{i})\\
    \frac{1}{N}\sum_{i}\frac{h(X_{i})D_{i}Z_{i}}{\hat{\tau}}&=\frac{1}{N_{1}}\sum_{Z_{i}=1}D_{i}h(X_{i})\\
    -\frac{1}{N}\sum_{i}\frac{h(X_{i})D_{i}(1-Z_{i})}{1-\hat{\tau}}&=-\frac{1}{N_{0}}\sum_{Z_{i}=0}D_{i}h(X_{i}),
\end{align*}
and it follows that 
\begin{equation*}
    \frac{1}{N}\sum_{i}\kappa_{i}h(X_{i})=\frac{1}{N}\sum_{i}h(X_{i})-\frac{1}{N_{1}}\sum_{Z_{i}=1}h(X_{i})+\frac{1}{N_{1}}\sum_{Z_{i}=1}D_{i}h(X_{i})-\frac{1}{N_{0}}\sum_{Z_{i}=0}D_{i}h(X_{i}).
\end{equation*}
With simple algebra, we can show that 
\begin{equation*}
    \frac{1}{N}\sum_{i}h(X_{i})-\frac{1}{N_{1}}\sum_{Z_{i}=1}h(X_{i})=(1-\hat{\tau})\left(\frac{1}{N_{0}}\sum_{Z_{i}=0}h(X_{i})-\frac{1}{N_{1}}\sum_{Z_{i}=1}h(X_{i})\right),
\end{equation*}
and the result for $\tilde{\chi}_{c}$ easily follows. \emph{QED}.

As a corollary to Proposition \ref{prop:plug-in-estimators}, expression (\ref{eq:sc_char_id_weight_wald}) is the estimand corresponding to the $\tilde{\chi}_{cc}$ estimator. Comparing (\ref{eq:sc_char_id_wald}) and (\ref{eq:sc_char_id_weight_wald}) reveals that the wedge between the two estimators, $\hat{\chi}_{cc}$ and $\tilde{\chi}_{cc}$, is due to the difference in the average characteristics between the treatment and control groups. While independence between $X$ and $Z$ implies this difference to be zero in the population, it is generally nonzero in a given sample. 

We can implement both $\hat{\chi}_{cc}$ and $\tilde{\chi}_{cc}$ using a standard 2SLS command in a statistical software. The only difference is that to implement the latter, we need to first transform $Y$ by subtracting from it the proportion of sample units assigned to the control group (i.e., $1-\hat{\tau}$). As it turns out, because $\hat{\tau}$ is consistent for $\tau$, we can proceed as if we know the true $\tau$, and the inference results reported by the 2SLS command are valid.

For our second claim, we remind the reader that the estimand in \citet{FinkNoto2019} is (the expression is provided in the Online Appendix F of their 2018 NBER working paper; note that the roles of $D$ and $Z$ are reversed in their notation):
\begin{equation}
\label{eq:FN_estimand}
    E[h(X)|D_{1}>D_{0}]=\frac{(s_{AT}+s_{C})\mu_{T}-s_{AT}\mu_{AT}}{s_{c}}
\end{equation}
where the various quantities are identified with
\begin{align*}
    s_{AT}&=\Pr(D=1|Z=0)\\
    s_{C}&=\Pr(D=1|Z=1)-\Pr(D=1|Z=0)\\
    \mu_{AT}&=E[h(X)|D=1,Z=0]\\
    \mu_{T}&=E[h(X)|D=1,Z=1].
\end{align*}
The resulting estimator, which we denote by $\hat{\chi}_{FN}$, is obtained by plugging in the sample analogs of $s_{AT}$, $s_{C}$, $\mu_{AT}$, and $\mu_{T}$. The following proposition establishes that $\hat{\chi}_{FN}$ can be alternatively implemented using a 2SLS regression with the product $h(X)D$ as the dependent variable, $D$ the endogenous variable, and $Z$ the instrument.

\begin{proposition}
\label{prop:FN_estimator}
The two estimators $\hat{\chi}_{FN}$ and $\hat{\chi}_{c}$ are equal. 
\end{proposition}
{\bf Proof of Proposition \ref{prop:FN_estimator}}: First notice that the estimand in (\ref{eq:FN_estimand}) is equivalent to 
\begin{equation}
\label{eq:FN_estimand_alt}
    E[h(X)|D_{1}>D_{0}]=\frac{\Pr(D=1|Z=1)\mu_{T}-\Pr(D=1|Z=0)\mu_{AT}}{\Pr(D=1|Z=1)-\Pr(D=1|Z=0)}.
\end{equation}
Since the denominator of (\ref{eq:FN_estimand_alt}) is the same as (\ref{eq:sc_char_id_wald}), we only need to show that the sample analogs of the numerators are equal. The sample analog of the first term in the numerator of (\ref{eq:FN_estimand_alt}) is
\begin{equation*}
\left(\frac{1}{N_{1}}\sum_{Z_{i}=1}D_{i}\right)\left(\frac{1}{\sum_{Z_{i}=1}D_{i}}\sum_{D_{i}=1,Z_{i}=1}h(X_{i})\right)=\frac{1}{N_{1}}\sum_{D_{i}=1,Z_{i}=1}h(X_{i})=\frac{1}{N_{1}}\sum_{Z_{i}=1}h(X_{i})D_{i},
\end{equation*}
which is the sample analog of $E[h(X)D|Z=1]$. Similarly, we can show that the second term $\Pr(D=1|Z=0)\mu_{AT}$ and  $E[h(X)D|Z=0]$ have the same sample analog. \emph{QED}.

The final claim of this section is that the relative asymptotic efficiency of the two estimators, $\hat{\chi}_{cc}$ and $\tilde{\chi}_{cc}$, depends on the underlying DGP. To see this, first focus on the numerators of the estimators. For simplicity, consider the case where $h$ is the identity function. Following standard arguments, the numerators of  $\hat{\chi}_{cc}$ and $\tilde{\chi}_{cc}$ have asymptotic variances
\begin{align*}
\text{avar of }\hat{\chi}_{cc}\text{ numerator}&:\text{ }\frac{1}{\tau}var(XY|Z=1)+\frac{1}{1-\tau}var(XY|Z=0)\\
\text{avar of }\tilde{\chi}_{cc}\text{ numerator}&:\text{ }\frac{1}{\tau}var(X(Y-1+\tau)|Z=1)+\frac{1}{1-\tau}var(X(Y-1+\tau)|Z=0).
\end{align*}
We further narrow in on the first term in each of the expressions above (the variance conditional on $Z=1$). Consider the simple (though likely unrealistic) case where $\tau=0.5$ and $X$ is binary, and where conditional on $Z=1$, $X$ and $Y$ are independent with $\Pr(X=1)=0.5$ and $\Pr(Y=1 | Z=1)=\mu_{Y}$. The independence allows us to express the two variances in terms of just the means and variances of $X$ and $Y$ conditional on $Z=1$. As a result, we can derive the difference between the two terms as
\begin{equation*}
    var(XY|Z=1)-var(X(Y-1+\tau)|Z=1)=0.25\mu_{Y}-0.0625
\end{equation*}
The implication is that in this simple setting, the first term of $\hat{\chi}_{cc}$'s numerator has a smaller variance than that of $\tilde{\chi}_{cc}$ if and only if $\mu_{Y}<0.25$. It is easy to construct similar scenarios under which one numerator or one estimator has an asymptotic variance that is larger or smaller than the other, so we omit the details here. For simplicity and practical consistency, we implement $\hat{\chi}_{cc}$ in this paper.

\subsection{Estimating Supercomplier Characteristics Quantiles\label{subsec_app:quantiles}}

\citet{Abadieetal2002} (henceforth, AAI) propose an estimation and inference procedure for the complier quantile treatment effect. They start from the observation that the $\theta$-quantile treatment effect parameter solves the population minimization problem
\begin{equation}
\label{eq:aai_quantile}
    (\alpha_{\theta},\beta_{\theta})=\argmin_{\alpha,\beta} E[\kappa\rho_{\theta}(Y-\alpha D-X\beta)],
\end{equation}
where $\rho_{\theta}(\lambda)\equiv(\theta-1_{[\lambda<0]})\lambda$ is the check function per \citet{BassettandKoenker1982}. AAI note that while the minimization problem (\ref{eq:aai_quantile}) is convex in the parameters, its sample analog is not because the $\kappa$ weight can be negative. Their solution comes from the insight that, by the law of iterated expectations, (\ref{eq:aai_quantile}) is equivalent to 
\begin{equation}
\label{eq:aai_implementation}
    (\alpha_{\theta},\beta_{\theta})=\argmin_{\alpha,\beta} E[\kappa_{\nu}\rho_{\theta}(Y-\alpha D-X\beta)],
\end{equation}
where $\kappa_{\nu}\equiv E[\kappa|U]$ with $U\equiv(D,Y,X)$. AAI show that $\kappa_{\nu}=\Pr(D_{1}>D_{0}|U)$ has a probability interpretation and is therefore nonnegative. They proceed to estimate the quantile effect by plugging in the estimated $\kappa_{\nu}$ in the sample analog of (\ref{eq:aai_implementation}). 

Our problem is analogous, but we need to adjust the AAI approach. The supercomplier characteristics $\theta$-quantile solves the problem
\begin{equation*}
    \gamma_\theta=\argmin_{\gamma}E[\pi\rho_{\theta}(X-\gamma)],
\end{equation*}
and we face the same challenge that $\pi$ may be negative. Like AAI, we can apply the law of iterated expectations and use a conditional version of $\pi$ instead. However, it turns out that when conditioning on the same triplet $U=(D,Y,X)$, $E[\pi|U]$ does not have a probability interpretation and may still be negative. In Lemma \ref{lemma:quantile} below, we show that we need to instead condition on just $V\equiv(Y,X)$ or simply $V \equiv X$. With the result in Lemma \ref{lemma:quantile}, we can adapt the estimation approach of AAI. Specifically, we can let $\pi_{\nu}\equiv E[\pi|V]$ and solve the sample analog of the minimization problem 
\begin{equation*}
    \gamma_{\theta}=\argmin_{\gamma}E[\pi_{\nu}\rho_{\theta}(X-\gamma)].
\end{equation*}
\begin{lemma}
\label{lemma:quantile}
Under Assumptions \ref{A:IV}.1-\ref{A:IV}.3 and \ref{A:OM}.1, 
\begin{equation*}
 E[\pi|V]=\Pr(G=cc|V).    
\end{equation*}
\end{lemma}
{\bf Proof of Lemma \ref{lemma:quantile}}: We examine the conditional expectation of each of the three terms that make up $\pi$: 
\begin{equation}
\label{eq:cond_pi}
    E[\pi|V]=E[\kappa|V]-E[\kappa_{0}Y|V]-E[\kappa_{1}(1-Y)|V]
\end{equation}
For the first term on the right hand side of (\ref{eq:cond_pi}), we can simply replace $U$ by $V$ in the proof of Lemma 3.2 by AAI and still have the analogous probability interpretation: $E[\kappa|V]=\Pr(D_{1}>D_{0}|V)$.

For the second term of (\ref{eq:cond_pi}),
\begin{equation*}
    E[\kappa_{0}Y|V]=E \left[\left. \frac{(1-D)(1-Z)Y}{\Pr(Z=0)}\right| V \right]-E\left[\left. \frac{(1-D)ZY}{\Pr(Z=1)}\right| V \right]
\end{equation*}
Because
\begin{align*}
&E[(1-D)(1-Z)Y|V]\\=&\Pr(D=0,Z=0,Y=1|V)\\=&\Pr(D_{0}=0,Z=0,Y_{0}=1|V)\\=&\sum_{k=0,1}\Pr(D_{0}=0,D_{1}=k,Z=0,Y_{0}=1|V)\\=&\sum_{k=0,1}\Pr(D_{0}=0,D_{1}=k,Y_{0}=1|V)\Pr(Z=0|D_{0}=0,D_{1}=k,Y_{0}=1,V)\\=&\Pr(G=na|V)\Pr(Z=0)+\Pr(G=ca|V)\Pr(Z=0),
\end{align*}
we have 
\begin{equation*}
E\left[\left.\frac{(1-D)(1-Z)Y}{\Pr(Z=0)}\right|V\right]=\Pr(G=na|V)+\Pr(G=ca|V).
\end{equation*}
At the same time, 
\begin{align*}
    &E[(1-D)ZY|V]\\
    =&\Pr(D=0,Z=1,Y=1|V)\\
    =&\Pr(D_{1}=0,Y_{0}=1|V)\Pr(Z=1|D_{1}=0,Y_{0}=1,V)\\
    =&\Pr(G=na|V)\Pr(Z=1),
\end{align*}
and therefore, 
\begin{equation*}
    E\left[\left.\frac{(1-D)ZY}{\Pr(Z=1)} \right| V\right]=\Pr(G=na|V).
\end{equation*}
It follows that 
\begin{equation*}
    E[\kappa_{0}Y|V]=\Pr(G=ca|V).
\end{equation*}
Similarly, we can show that the third term of (\ref{eq:cond_pi}) is
\begin{equation*}
    E[\kappa_{1}(1-Y)|V]=\Pr(G=cn|V).
\end{equation*}
Since the three groups $G=ca,cn,cc$ partition the set of treatment compliers, we prove Lemma \ref{lemma:quantile} by putting the three terms together. \emph{QED.}

\subsection{Implications under Conditional Independence\label{subsec_app:cond_indep}}

In this section, we analyze the supercomplier characteristics estimand under the conditional independence assumption. Specifically, we investigate identification under the corresponding 2SLS regression of (\ref{eq:stata}) with covariate controls, and we focus on the case of stratified experiments where the controls are the set of stratum dummies. We show that the population 2SLS estimator identifies a non-negatively weighted average of supercomplier characteristics, as opposed to an expression where other unobserved subpopulations receive positive weight. This result is analogous to the \citet{Blandholetal2022} finding on ``weak causality"---under certain conditions, the LATE estimand with covariate controls identifies a non-negatively weighted average of complier treatment effects.

\begin{assumption}[Stratified Experiment] \label{A:stratified}
\leavevmode
\begin{enumerate}[label={\arabic*.}, align=left]
\item Stratified Random Assignment: $(Y_{00}, Y_{01}, Y_{10}, Y_{11}, D_0, D_1) \indep Z$ conditional on $W$ and $0<\Pr(Z=1|W)<1$.
\item Conditional Reduced Form: The population coefficient on $Z$ from the reduced form regression of $Y$ on $Z$ and $W$ is nonzero.
\item Saturation: $E[Z | W] = \LP(Z | W)$, where $\LP(Z | W)$ is the linear projection of $Z$ on $W$.
\end{enumerate}
\end{assumption}

Assumptions \ref{A:stratified}.1 and \ref{A:stratified}.2 are the covariate-control counterparts to Assumptions \ref{A:IV}.1 and \ref{A:OM}.2, respectively. Assumption \ref{A:stratified}.3 stipulates that the true conditional probability of treatment eligibility assignment can be recovered with linear regression. This condition is satisfied when the controls are saturated stratum fixed effects in a stratified randomized experiment.

\begin{proposition}
\label{prop:SC_Weights}
Denote the population coefficient from a supercomplier characteristics regression with covariate control $W$ by $\beta_{2SLS}$. Under Assumptions \ref{A:IV}.2, \ref{A:IV}.3, \ref{A:OM}.2 and 3, and provided that $X \indep Z | W$, 
\begin{equation}
    \beta_{2SLS}=E\big[\omega_{W}E[h(X)|G=cc,W]\big],
\end{equation}
where the weight $\omega_{W}$ is nonnegative for all values of $W$.
\end{proposition}

{\bf Proof of Proposition \ref{prop:SC_Weights}}: We define $\tilde{Z} = Z - \LP(Z | W)$ and use Proposition 6 of \citet{Blandholetal2022} to write
\begin{equation*}
\beta_{2SLS} =  \frac{E[h(X)Y \tilde{Z}]}{E[Y \tilde{Z}]}.
\end{equation*}
Decompose the numerator of $\beta_{2SLS}$ as
\begin{align}
\label{eq:BBMT_decomp}
E[h(X)Y\tilde{Z}]&=E[E[h(X)Y\tilde{Z}|W]]\nonumber\\
&=E\big[E[cov(h(X)Y,\tilde{Z}|W)]+E[E[h(X)Y|W]E[\tilde{Z}|W]\big]\nonumber\\
&=E\big[\underbrace{E[cov(h(X)Y,Z|W)]}_{\text{(i)}}+\underbrace{E[E[h(X)Y|W]E[\tilde{Z}|W]]}_{\text{(ii)}}\big],
 \end{align}
where the last equality follows from the fact that conditional on $W$, $\LP(Z | W)$ is a constant and is therefore uncorrelated with $h(X)Y$.

We can further decompose term (i) of equation (\ref{eq:BBMT_decomp}) as:
\begin{align}
\label{eq:BBMT_decomp_term_i}
cov(h(X)Y,Z|W)&=E[h(X)Y(Z-E[Z|W])|W]\nonumber\\
&=E[h(X)Y|Z=1,W](1-E[Z|W])E[Z|W]\nonumber\\
&\;\;\;-E[h(X)Y|Z=0,W](E[Z|W])(1-E[Z|W])\nonumber\\
&=(E[h(X)Y|Z=1,W]-E[h(X)Y|Z=0,W])(1-E[Z|W])E[Z|W]\nonumber\\
&=E[h(X)|G=cc,W]\times RF_{W}\times(1-E[Z|W])E[Z|W]
\end{align}
where the last equality follows from the conditional variation of Proposition \ref{prop:SC_Char} with $RF_W \equiv E[Y | Z=1,W] - E[Y | Z=0,W]$ being the reduced form conditional on $W$.

For term (ii) of equation (\ref{eq:BBMT_decomp}), note that 
\begin{equation*}
 E[\tilde{Z} | W ]  = E[Z | W] - \LP(Z | W) = 0   
\end{equation*}
if $E[Z | W] = \LP(Z | W)$. Thus, term (ii) is zero whenever the expectation of $Z$ in linear in $W$, which is true in stratified experiments with $W$ being the full set of stratum dummies. 

Putting together equations (\ref{eq:BBMT_decomp_term_i}) and (\ref{eq:BBMT_decomp_term_ii}), we have that the numerator of $\beta_{2SLS}$ is 
\begin{equation}
\label{eq:BBMT_decomp_term_ii}
    E[h(X)Y\tilde{Z}]=E\big[E[h(X)|G=cc,W]\times RF_{W}\times(1-E[Z|W])E[Z|W]\big].
\end{equation}
We can simiarly show that the denominator of $\beta_{2SLS}$ is:
\begin{equation*}
E[h(X)Y\tilde{Z}]=E\big[RF_{W}\times(1-E[Z|W])E[Z|W]\big],
\end{equation*}
and the result of Proposition \ref{prop:SC_Weights} follows with weight $\omega_W=RF_{W}\times(1-E[Z|W])E[Z|W]$. Since $RF_W \geq 0$ by Assumption \ref{A:stratified}.2 and $Z$ is binary, $\omega_W \geq 0$ for all values of $W$. \emph{QED.}

\subsection{Joint Test of Identifying Assumptions for a General $Y$}\label{subsec:generalized_test}
The formulation of the joint test of identifying assumptions in Proposition \ref{prop:sharp} requires the outcome variable $Y$ be binary. In this section, we provide a generalized test that does not impose any restriction on the distribution of $Y$. Our arguments follow those of \citet{Kitagawa2015} but incorporate outcome monotonicity of Assumption \ref{A:OM}. 

\begin{proposition}
\label{prop:generalized_test}
Define the conditional probability distributions
\begin{align*}
P(B,d) & \equiv\Pr(Y\in B,D=d|Z=1)\\
Q(B,d) & \equiv\Pr(Y\in B,D=d|Z=0),
\end{align*}
where B is a Borel set of $\mathbb{R}$. 
(i) Under Assumptions \ref{A:IV}.1-\ref{A:IV}.3 and \ref{A:OM}.1, these inequalities hold:
\begin{align}
P(B,1)-Q(B,1) & \geq0\label{eq:generalized_test_1}\\
Q(B,0)-P(B,0) & \geq0\label{eq:generalized_test_2}\\
Q(A,1)-P(A,1) & \geq P(A,0)-Q(A,0),\label{eq:generalized_test_3}
\end{align}
where $B$ is any Borel set and $A$ is any lower Borel interval of the form $A=(-\infty,y]$.
(ii) If $P$ and $Q$ satisfy inequalities (\ref{eq:generalized_test_1})-(\ref{eq:generalized_test_3}), then there exists a joint probability law
$(Y_{11},Y_{10},Y_{01},Y_{00},D_{1},D_{0},Z)$ that satisfies Assumptions \ref{A:IV}.1-\ref{A:IV}.3 and \ref{A:OM}.1 and induces $P$ and $Q$.
\end{proposition}

\begin{proof}
(i) For this part, \citet{Kitagawa2015} already shows that Assumptions \ref{A:IV}.1-\ref{A:IV}.3 imply (\ref{eq:generalized_test_1}) and (\ref{eq:generalized_test_2}). We only need to show that our identifying assumptions imply (\ref{eq:generalized_test_3}), or equivalently, for any $y \in \mathbb{R}$:
\begin{equation}
    \Pr(Y \leq y|Z=1) \leq \Pr(Y \leq y|Z=0).\label{eq:stochastic_dominance}
\end{equation}
Inequality (\ref{eq:stochastic_dominance}) holds under Assumptions \ref{A:IV}.1-\ref{A:IV}.3 and \ref{A:OM}.1 because 1) the $Y$ distribution is the same among always takers and the same among never takers across the treatment and control groups and 2) the $Y$ distribution among treated compliers first-order stochastically dominates the $Y$ distribution among untreated compliers:\footnote{Always takers, never takers, compliers, and defiers in this proof refer to the treatment always takers, treatment never takers, treatment compliers, and treatment defiers.}
\begin{align*}
&\Pr(Y\leq y|Z=1)\\
= & \Pr(Y\leq y|Z=1,D_{1}=D_{0})\Pr(D_{1}=D_{0}|Z=1)+\\
&\Pr(Y_{1}\leq y|Z=1,D_{1}>D_{0})\Pr(D_{1}>D_{0}|Z=1)\\
\leq	& \Pr(Y\leq y|Z=0,D_{1}=D_{0})\Pr(D_{1}=D_{0}|Z=0)+\\
&\Pr(Y_{0}\leq y|Z=0,D_{1}>D_{0})\Pr(D_{1}>D_{0}|Z=0)\\
= &	\Pr(Y\leq y|Z=0).
\end{align*}
(ii) We prove this part of the proposition by slightly modifying the construction of \citet{Kitagawa2015} in the proof of his Proposition 1.1. \citet{Kitagawa2015} assumes instrument exclusion (Assumption \ref{A:IV}.2) and constructs a joint distribution of $(Y_{1},Y_{0},D_{1},D_{0},Z)$, which satisfies random assignment and treatment monotonicity (Assumptions \ref{A:IV}.1 and \ref{A:IV}.3), and which induces the observed $P$ and $Q$ satisfying the first two inequalities (\ref{eq:generalized_test_1}) and (\ref{eq:generalized_test_2}).

We proceed similarly, with the difference being that the joint distribution also satisfies outcome monotonicity (Assumption \ref{A:OM}.1) and that the induced $P$ and $Q$ satisfy all three inequalities (\ref{eq:generalized_test_1})-(\ref{eq:generalized_test_3}). Our modification of \citet{Kitagawa2015}'s construction concerns the unobserved distributions of $Y_{0}$ among always takers and of $Y_{1}$ among never takers. Specifically, \citet{Kitagawa2015} lets $\Pr(Y_{0}\in B|\text{always taker})$ and $\Pr(Y_{1}\in B|\text{never taker})$---the unobserved distribution of $Y_{0}$ among always takers and the unobserved distribution of $Y_{1}$ among never takers respectively---be arbitrary as long as $\Pr(Y_{0}\in\mathbb{R}|\text{always taker})=\Pr(D=1|Z=0)$ and $\Pr(Y_{1}\in\mathbb{R}|\text{never taker})=\Pr(D=0|Z=1)$. Because of our outcome monotonicity assumption, we additionally require $Y_{1}$ to have weak first-order stochastic dominance over $Y_{0}$ within each of the two populations. For ease of exposition, we use the simple construction where $Y_{1}$ and $Y_{0}$ have the
same distribution among always takers and among never takers, respectively. That is, $\Pr(Y_{0}\in B_{0},D=1|Z=0)=Q(B_{0},1)$ and $\Pr(Y_{1}\in B_{1},D=0|Z=1)=P(B_{1},0)$ for all Borel sets $B_{0}$ and $B_{1}$. 

The rest of the construction of a joint distribution of $(Y_{1},Y_{0},D_{1},D_{0},Z)$ is the same as that in \citet{Kitagawa2015}. For any Borel sets $B_{1}$ and $B_{0}$,
\begin{align}
 & \Pr(Y_{1}\in B_{1},Y_{0}\in B_{0},D_{1}=1,D_{0}=0|Z=1)\nonumber\\
= & \Pr(Y_{1}\in B_{1},Y_{0}\in B_{0},D_{1}=1,D_{0}=0|Z=0)\nonumber\\
\equiv & \frac{P(B_{1},1)-Q(B_{1},1)}{P(\mathbb{R},1)-Q(\mathbb{R},1)}\cdot\frac{Q(B_{0},0)-P(B_{0},0)}{Q(\mathbb{R},0)-P(\mathbb{R},0)}\cdot\nonumber\\
& \left[P(\mathbb{R},1)-Q(\mathbb{R},1)\right]\cdot1_{[P(\mathbb{R},1)-Q(\mathbb{R},1)>0]}\label{eq:Kitagawa}
\end{align}
\begin{align*}
 & \Pr(Y_{1}\in B_{1},Y_{0}\in B_{0},D_{1}=0,D_{0}=0|Z=1)\\
= & \Pr(Y_{1}\in B_{1},Y_{0}\in B_{0},D_{1}=0,D_{0}=0|Z=0)\\
\equiv & \frac{P(B_{1},0)}{P(\mathbb{R},0)}\cdot\frac{P(B_{0},0)}{P(\mathbb{R},0)}\cdot P(\mathbb{R},0)\cdot1_{[P(\mathbb{R},0)>0]}
\end{align*}
\begin{align*}
 & \Pr(Y_{1}\in B_{1},Y_{0}\in B_{0},D_{1}=1,D_{0}=1|Z=1)\\
= & \Pr(Y_{1}\in B_{1},Y_{0}\in B_{0},D_{1}=1,D_{0}=1|Z=0)\\
\equiv & \frac{Q(B_{1},1)}{Q(\mathbb{R},1)}\cdot\frac{Q(B_{0},1)}{Q(\mathbb{R},1)}\cdot Q(\mathbb{R},1)\cdot1_{[Q(\mathbb{R},1)>0]}
\end{align*}
\begin{align*}
 & \Pr(Y_{1}\in B_{1},Y_{0}\in B_{0},D_{1}=0,D_{0}=1|Z=1)\\
= & \Pr(Y_{1}\in B_{1},Y_{0}\in B_{0},D_{1}=0,D_{0}=1|Z=0)\\
\equiv & 0,
\end{align*}
where a product is defined to be zero if the indicator function inside equals zero. Following the arguments in \citet{Kitagawa2015}, this constructed joint distribution is a probability measure, satisfies Assumptions \ref{A:IV}.1 and \ref{A:IV}.3, and induces the observed distributions of $P$ and $Q$. 

We just need to show that the joint distribution also satisfies outcome monotonicity. Since outcome monotonicity holds for the always taker, never taker, and defier populations by construction, we only need to examine the compliers.
For any lower Borel interval $A$, 
\begin{align*}
 & \Pr(Y_{1}\in A,D_{1}=1,D_{0}=0)\\
= & P(A,1)-Q(A,1)\\
\leq & Q(A,0)-P(A,0)\\
= & \Pr(Y_{0}\in A,D_{1}=1,D_{0}=0).
\end{align*}
We obtain the first (last) equality by replacing $B_1$ with $A$ and $B_0$ with $\mathbbm{R}$ ($B_1$ with $\mathbbm{R}$ and $B_1$ with $A$) in (\ref{eq:Kitagawa}), and by using the identities of $P(\mathbbm{R},0)=1-P(\mathbbm{R},1)$ and $Q(\mathbbm{R},0)=1-Q(\mathbbm{R},1)$ based on the definitions of $P$ and $Q$. The inequality in the middle is simply (\ref{eq:generalized_test_3}), which is assumed to hold. \textit{QED}.
\end{proof}

In the specific case of a binary $Y$, we already have a simple test as discussed in Section \ref{subsec:OM}. To implement the joint test based on Proposition \ref{prop:generalized_test} for a non-binary $Y$, we note that the first two inequalities (\ref{eq:generalized_test_1}) and (\ref{eq:generalized_test_2}) are exactly the same as those that underlie the tests by \citet{Kitagawa2015} and \citet{MourifieandWan2017}. \citet{Kitagawa2015} proposes a variance-weighted Kolmogorov-Smirnov test statistic for these inequalities and provides an implementation in R. Meanwhile, \citet{MourifieandWan2017} transform (\ref{eq:generalized_test_1}) and (\ref{eq:generalized_test_2}) into moment inequalities where the moments are conditional on the value of $Y$, allowing their test to be implemented using a Stata command by \citet{Chernozhukovetal2015} based on  \citet{ChernozhukovLeeRosen2013}. Since most of our intended audience uses Stata, we build on \citet{MourifieandWan2017} for convenience, though R users could easily implement a similar joint test based on the \citet{Kitagawa2015} procedure. 

Because inequality (\ref{eq:generalized_test_3}) only holds for lower Borel intervals and not just any Borel set, it cannot be turned into the same type of moment condition and incorporated into the \citet{Chernozhukovetal2015} command. However, it is straightforward to test (\ref{eq:generalized_test_3}) with a standard one-sided two-sample Kolmogorov-Smirnov test via the \texttt{ksmirnov} command in Stata. Consequently, our procedure entails testing (\ref{eq:generalized_test_1}) and (\ref{eq:generalized_test_2}) separately from (\ref{eq:generalized_test_3}), and we will conclude that one or more of our identifying assumptions are violated if either test rejects the null at the 5 percent level. 

It is well known that conducting two tests separately like this generally fails to control the overall size, and a common remedy is to apply a multiple hypothesis correction like Bonferroni. However, a Bonferroni procedure is conservative, which can be a virtue in treatment effect testing but a vice in assumption or specification testing---a conservative test tends to lack power, which means that we would often fail to reject incorrect assumptions. Therefore, we choose not to correct for multiple hypotheses and note that the actual (asymptotic) size of our procedure is between 5 and 10 percent. In our empirical applications, we use the test from Section \ref{subsec:OM} (which controls size at 5 percent) for binary outcomes and the test described here for continuous outcomes.

\section{Testing Outcome Monotonicity in \citet{Bitleretal2006,Bitleretal2017}}\label{sec_app:outcome_monotonicity_CTJF}

As a proof of concept, we implement our outcome monotonicity test on the Jobs First evaluation data. Briefly, the Jobs First evaluation, implemented by the Manpower Demonstration Research Corporation (MDRC) beginning in 1997, was an experimental evaluation of Connecticut's Jobs First welfare reform initiative (see \citealt{Bloometal2002} for a description). The evaluation compared outcomes like earnings, employment, and welfare use among individuals randomly assigned to the Jobs First program to those in the traditional Connecticut welfare program. The Jobs First program imposed a time limit on welfare receipt and allowed participants to earn up to the federal poverty line without any decrease to their benefits. 

As discussed in \citet{Bitleretal2006,Bitleretal2017}, based on theory of labor supply, the predicted effect of the Jobs First earnings disregard relative to the traditional welfare benefit schedule should vary across the earnings distribution. For women with previously low earnings, the intervention is predicted to increase earnings. However, for women with relatively high earnings, the intervention could have a negative effect. Thus, Jobs First is an experiment where we would expect the outcome monotonicity assumption to fail and where it has indeed been shown to fail by Figure 3 of \citet{Bitleretal2006}. It therefore provides a useful test case for our proposed outcome monotonicty test.  

To examine the viability of our test, we use data by \citet{Bloom2022Data} to assess whether the Jobs First treatment had a monotonically positive effect on participants' average quarterly earnings in the first seven quarters after random assignment. In the experiment, there is no non-compliance, so we limit our focus to inequality (\ref{prop_sharp_ineq:3}). Our approach follows the subsample analyses in \citet{Bitleretal2017}. We split participants into three groups, $r=\{1,2,3\}$, based on their earnings seven quarters prior to random assignment---no earnings, earnings below median, and earnings at or above median earnings---where the median is measured only among participants with non-zero earnings. 

We then create three copies of the data and estimate the following stacked regression:
\begin{align*}
    Y_i^r = \phi^r + \theta_1 Z_i 1_{[r=1]} + \theta_2 Z_i 1_{[r=2]} + \theta_3 Z_i 1_{[r=3]} + \varepsilon_i^r
\end{align*}
Similar to equation (\ref{eq:joint_test}), $r$ indexes each stack, $\phi^r$ is the stack specific constant, and standard errors are clustered at the individual level. However, in this case, $Y_i^r$ is simply equal to average quarterly earnings in the first seven quarters of Jobs First for all stacks. Note that the choice of three stacks here corresponds to the number of subsamples examined, not to the inequalities in Proposition \ref{prop:sharp}.

We test whether $\theta\equiv \min(\theta_1,\theta_2,\theta_3)$ is nonnegative by taking draws from a joint normal distribution with the estimated covariance matrix from the stacked regression. Our estimate of $\theta$ is -347.257 while the 5-percent critical value from the simulated normal distribution is -303.591. Thus, the test rejects outcome monotonicity at the 5-percent level.

We note, however, that our ability to reject outcome monotonicity varies based on the subsamples we examine. This pattern is similar to the results in \citet{Bitleretal2017}, where the authors find that using alternative subsamples does not adequately reflect treatment effect heterogeneity. We performed similar tests with all of the subsamples in Table 1 of \citet{Bitleretal2017}, which includes subsamples by education of the participant, by whether the youngest child in the case is under 6 years old, by number of children in the case, by marital status of the participant, by number of quarters with any earnings before random assignment, and by whether the participant was receiving welfare benefits seven quarters before random assignment. We cannot reject outcome monotonicity at standard significance levels in any of these tests. 

We conclude from this proof of concept that our proposed outcome monotonicity test can detect the presence of subsamples for which the outcome monotonicity assumption fails. However, the choice of subsamples to examine matters.

\section{Weighted MVPF and Supercompliers\label{sec_app:sc_weight_cts}}
In this section, we show that weighted willingness to pay (the denominator of MVPF) equals unweighted WTP (the local average treatment effect) multiplied by our identified supercomplier characteristics for any distribution of the outcome variable. Our result here nests equation (\ref{eq:sc_wt}) as a special case for binary $Y$. 

The weighted willingness to pay for the complier group is
\begin{align*}
 & E[\eta_{i}(Y_{1i}-Y_{0i})|\text{complier}]\\
= & E[\eta_{i}(Y_{1i}-Y_{0i})|\text{complier},Y_{1i}>Y_{0i}]\Pr(Y_{1i}-Y_{0i}>0|\text{complier})\\
= & E[\eta_{i}(Y_{1i}-Y_{0i})|\text{supercomplier}]\Pr(Y_{1i}-Y_{0i}>0|\text{complier})\\
= & \frac{E[\eta_{i}Y_{i}|Z_{i}=1]-E[\eta_{i}Y_{i}|Z_{i}=0]}{E[Y_{i}|Z_{i}=1]-E[Y_{i}|Z_{i}=0]}\cdot E[Y_{1i}-Y_{0i}|\text{supercomplier}]\Pr(Y_{1i}-Y_{0i}>0|\text{complier})\\
= & \frac{E[\eta_{i}Y_{i}|Z_{i}=1]-E[\eta_{i}Y_{i}|Z_{i}=0]}{E[Y_{i}|Z_{i}=1]-E[Y_{i}|Z_{i}=0]}\cdot E[Y_{1i}-Y_{0i}|\text{complier},Y_{1i}>Y_{0i}]\Pr(Y_{1i}-Y_{0i}>0|\text{complier})\\
= & \frac{E[\eta_{i}Y_{i}|Z_{i}=1]-E[\eta_{i}Y_{i}|Z_{i}=0]}{E[Y_{i}|Z_{i}=1]-E[Y_{i}|Z_{i}=0]}\cdot E[Y_{1i}-Y_{0i}|\text{complier}]\\
= & \frac{E[\eta_{i}Y_{i}|Z_{i}=1]-E[\eta_{i}Y_{i}|Z_{i}=0]}{E[Y_{i}|Z_{i}=1]-E[Y_{i}|Z_{i}=0]}\text{LATE}_{Y}.
\end{align*}
It follows by Proposition \ref{prop:non_binary_Y} that the first term represents the supercomplier average of the weights under our identifying assumptions.

\end{document}